\documentclass[aps,pre,reprint,groupedaddress,longbibliography,floatfix]{revtex4-2}

\usepackage{amsmath}
\usepackage{amssymb}
\usepackage{graphicx,psfrag}
\usepackage{verbatim}
\usepackage{color}
\usepackage{algorithm}
\usepackage{algpseudocode}

\graphicspath{{/home/noah/Documents/LaTeX Projects/Integer_Lattice_Gas_with_a_sampling_collision_operator_for_the_fluctuating_Navier_Stokes_Equation/}}

\begin{document}


\title{Integer Lattice Gas with a sampling collision operator for the fluctuating Navier-Stokes Equation}


\author{Noah Seekins}
\email{noah.seekins@ndsu.edu}
\author{Alexander J. Wagner}
\email[]{alexander.wagner@ndsu.edu}
\homepage[]{www.ndsu.edu/pubweb/$\sim$carswagn}
\affiliation{Department of Physics, North Dakota State University, Fargo, North Dakota 58108, USA}


\date{\today}

\begin{abstract}
  This paper constitutes a step in the direction of developing integer lattice gas methods as an attractive alternative to lattice Boltzmann methods. Here we show that to Boltzmann limit the one dimensional Blommel integer lattice gas is very close to entropic lattice Boltzmann. More interestingly the integer lattice gas retains additional correlations that prevent the existence of a well defined Boltzmann limit. In the analysis of the decaying sine wave we will see that in some situations the bulk viscosity can crucially depend on such correlations beyond the Boltzmann limit. A sampling collision operator, introduced here, can speed up the execution time to make the algorithm obtain comparable computational efficiency to entropic lattice Boltzmann methods.
\end{abstract}


\maketitle

\section{Introduction}

Lattice gas methods for simulating hydrodynamics have a long history, but they have fallen out of favor compared to lattice Boltzmann methods which were originally derived from lattice gas methods. We will give a brief history of this development in our introduction, and then argue that novel integer lattice gas approaches may both be competitive with lattice Boltzmann methods and allow for additional correlations and fluctuations lost in the Boltzmann description to be retained.

Frisch, Hasslacher and Pomeau's \cite{frisch1987lattice} extension of lattice gas automata to allow for the simulation of hydrodynamics opened up the new field of discrete simulations of fluid dynamics. Despite their very simple local evolution rules these lattice gases could be shown to recover the continuity and Navier-Stokes equations. The fluctuations of these lattice gases closely resembled the fluctuations of real liquids \cite{Rivet_Boon_2001}.

Initially the Boltzmann limit was applied to lattice gas systems to analyze their hydrodynamic limit in analogy to the procedure in kinetic theory~\cite{Huang1987Statistical}. This Boltzmann limit consists of taking an ensemble average that removes the fluctuations. McNamara \textit{et al.}~\cite{mcnamara1988use} were the first to use this Boltzmann limit of the lattice gas as a simulation method in its own right. Because simulating the lattice Boltzmann equation eliminated statistical noise, this was an advantage for simulating noise-free hydrodynamics. Previously this had to be achieved by performing a statistical average over several lattice gas simulations.

The lattice Boltzmann method was later decoupled from the lattice gas entirely when Qian, d'Humieres, and Lallemand~\cite{qian1992lattice} proposed a simplified BGK collision operator. By simply choosing a local equilibrium distribution (the now ubiquitous second order polynomial equilibrium distribution) they simplified the algorithm and made it significantly more accessible.  They were also able to remove some non-Galilean invariant artifacts of Boolean lattice gas systems, such as a velocity dependent viscosity~\cite{frisch1987lattice, higuera1989boltzmann} and pressure, even if these deviations became most notable in parameter ranges (i.e. at high Mach numbers) that were still not accessible to these lattice Boltzmann methods.

The simplicity of this BGK collision operator, more than anything else, influenced the explosive rise of lattice Boltzmann methods over their ancestral lattice gas methods. Certain systems, however, such as the Brownian motion of colloids considered by Ladd \textit{et al.}~\cite{ladd1988application}, require fluctuations to be simulated. These fluctuating systems were, at the time, limited to being simulated by lattice gases. To allow these simulations to benefit from the advances of lattice Boltzmann methods, Ladd developed a fluctuating lattice Boltzmann method. The basic idea was to implement Landau and Lifshitz' \cite{Landau1959fm} fluctuating stress tensor into the collision operator leading to a Langevin approach~\cite{langevin} for the fluctuating lattice Boltzmann method. This fluctuating lattice Boltzmann method recovered the fluctuating hydrodynamic equations~\cite{ladd1993short}. However, utilizing the fluctuating stress tensor alone limited the recovery of the correct fluctuations to those with a wavelength on the order of the system size, while the total order of fluctuations was suppressed by about a factor of two.

Adhikari \textit{et al.} discovered that to recover the smaller scale fluctuations, a full set of fluctuating modes, rather than just those associated with hydrodynamic quantities, was needed~\cite{adhikari2005fluctuating}. This method is restricted to ideal gas systems. For  non-ideal systems additional corrections are required~\cite{belardinelli2015fluctuating}. Since the Langevin approach is based on continuous densities, the large fluctuations at low densities will lead to the formation of non-physical negative densities~\cite[sec. 5.3]{Kahler2012Thesis}.

When considering methods for simulating fluctuating systems, however, it is much more natural to think in terms of discrete systems since fluctuations fundamentally arise because of the discreteness of nature \cite{einstein1906theorie}. This inspired Chopard \textit{et al.} to instead develop novel lattice gas algorithms instead of adding fluctuations artificially to the lattice Boltzmann algorithm~\cite[sec. 5.7]{chopard1998multiparticle, chopardcellular}. They extended the occupation numbers of the lattice gas from Boolean values to integer values, a concept that was first introduced by Chen \textit{et al.} the year prior with their Digital Physics approach~\cite{chen1997digital}. This allowed for the lattice gas system of Chopard \textit{et al.} to recover a Poisson distribution in global equilibrium. Secondly, Chopard \textit{et al.} implemented a sampling algorithm, which was originally theorized by Boghosian \textit{et al.}~\cite{boghosian1997integer}, that distributed particles randomly across the system velocities based on a local equilibrium ensemble. For the diffusive case, a multinomial distribution over the occupation numbers themselves was utilized for the local equilibrium ensemble, which was sufficient to follow the conservation of mass required by the system~\cite[Sec. 5.7]{chopardcellular}. 

In extending the method to hydrodynamic systems, however, Chopard \textit{et al.} did not utilize an explicit local equilibrium ensemble due to the complexity involved in deriving this distribution while accounting for momentum conservation. Instead, a multinomial distribution similar to that of the diffusive case was used, centered around the second order polynomial equilibrium distribution. This on its own would violate momentum conservation, so a random factor that redistributed a single particle at a time until momentum conservation was upheld was also added~\cite{chopard1998multiparticle}. Both of these methods were, for higher densities, well approximated by Gaussian distributions, which caused their algorithmic complexity to be approximately O(1) with respect to the system density, matching the complexity of the lattice Boltzmann algorithm. 

Blommel and Wagner created another integer based lattice gas algorithm to simulate hydrodynamic systems~\cite{blommel2018integer}. Their algorithm utilized a series of random two-particle collisions to redistribute particle velocities while still maintaining momentum conservation with collision rules explicitly derived by imposing detailed balance on the system. This allowed Blommel and Wagner to recover Poisson distributed fluctuations in global equilibrium similar to Chopard \textit{et al.}.

When analyzing their algorithm, the global equilibrium average distribution was found to correspond not to the polynomial lattice Boltzmann equilibrium distribution, but to the entropic lattice Boltzmann equilibrium distribution. This equilibrium distribution had already been derived by Ansumali \textit{et al.} \cite{Ansumali_2003} previously by minimizing an H-functional.  The polynomial equilibrium distribution utilized by Chopard \textit{et al.} \cite{chopard1998multiparticle} for their hydrodynamic multiparticle lattice gas, therefore, was actually incompatible with a microscopic lattice gas implementation. 

Since the number of individual two-particle collisions needed to achieve a state close to local equilibrium scales with the square of the number of particles at each lattice site, Blommel and Wagner's algorithm proved much slower than the fluctuating lattice Boltzmann algorithm even for moderate local densities of a few hundred particles per lattice site. This lead Seekins and Wagner~\cite{seekins2022integer} to accelerate Blommel's algorithm. A sampling method similar to that of Chopard \textit{et al.} was implemented to reduce the number of random numbers needed, leading to increased efficiency. 

Using detailed balance to derive a local equilibrium distribution, Seekins and Wagner recovered an identical multinomial distribution to that of Chopard \textit{et al.}~\cite[Sec. 5.7]{chopardcellular}. Although this meant that the diffusive algorithm created by Seekins and Wagner was essentially identical to that of Chopard \textit{et al.}, the derivation of the equilibrium distribution via detailed balance was novel, and we show here that it can be used to derive the proper equilibrium distribution for hydrodynamic systems. The efficiency of the algorithm was comparable to a diffusive fluctuating lattice Boltzmann algorithm, along with having a complexity at high densities of approximately O(1) with respect to density~\cite{seekins2022integer, wagner2016fluctuating}.

In this paper, we extend our prior work for diffusive systems to hydrodynamic systems by deriving an appropriate local equilibrium distribution. For simplicity, we limit the system to one dimension, leaving an extension to higher dimensions to future work. The paper is structured as follows: in Section \ref{sec: Sampling LG} of the paper, we will cover the sampling lattice gas algorithm in detail and derive the local equilibrium distribution and provide an analysis of the Poisson distribution as a global equilibrium distribution for the integer lattice gas. Section III will discuss the Boltzmann limit of the lattice gas and its uniqueness. We show that there isn't a unique Boltzmann limit, although for many practical situations the BGK collision operator can be a good approximation for the Boltzmann limit. Section IV covers the results of two test cases used to study the non-equilibrium dynamics of the system. Mostly we demonstrate excellent agreement between the averaged lattice gas and the equivalent lattice Boltzmann approach. However, in certain situations there is clear evidence of lattice gas behavior beyond the Boltzmann limit. Finally, Section V will compare the efficiency of our sampling lattice gas to that of both Blommel and Wagner and an entropic lattice Boltzmann algorithm showing that the new implementation of the integer lattice gas shows significant speed up to make it comparable in efficiency with the entropic lattice Boltzmann method.


\section{The Sampling Lattice Gas Algorithm\label{sec: Sampling LG}}
\subsection{Sampling Lattice Gas Basics}
A lattice gas consists of a set of discrete cells situated on a lattice, each containing a discrete number of velocity states $v_i$ for particles to occupy~\cite{wolf2004lattice}. The velocities are usually identical for every cell, and are defined such that if $x$ is a position on the lattice, then for any velocity $v_i$, $x+v_i$ is also a position on the lattice, assuming a timestep of $\Delta t=1$. These velocities are also typically restricted in number such that the particle displacement over a single timestep is small. The particles are accounted for by the occupation numbers $n_i(x,t)$, denoting the particles that move from lattice site $x-v_i$ at time $t-1$ to lattice site $x$ at time $t$. These particles are then redistributed among the occupation numbers, observing the applicable conservation rules, and streamed. This two-step process represents a discrete timestep for the system. Thus, the $n_i(x, t)$ evolve according to,
\begin{equation}
    n_i(x+v_i, t+1)=n_i(x, t)+\Xi_i[\{n_j\}_g],
    \label{eqn: lgEvo}
\end{equation}
which is known as the lattice gas evolution equation. Here $\{n_j\}_g$ represents the set of all $n_i$ of the system at the current time, and $\{n_j\}_l$ represents the set of $n_i$ at a single lattice site at the current time, with the subscripts $l$ and $g$ denoting between the two. In this evolution equation, the collision operator $\Xi_i$ is a stochastic operator. The explicit nature of the $n_i$ is dependent on the lattice gas system observed. For Boolean lattice gases, for example, these $n_i$ would be Boolean in nature, leading to $\Xi_i$ to be calculable explicitly within a lookup table due to the limited nature of the probability space for single-occupation velocity states~\cite{FHP, frisch1987lattice}. In our integer lattice gas method, the $n_i$ values become integer based, which leads to a collision operator $\Xi_i$ that can take on integer values.

In the ideal gas systems considered here the collision operator will only act on the local distribution of occupation numbers, $\{n_j\}_l$, which consists only of the local occupation numbers at position $x$ and $t$. In this nomenclature we exclude the implicit dependence on $x$ and $t$ to make the equations more readable. However, in principle the collision operator can depend on the global distribution $\{n_j\}_g$, as would be the case for non-ideal lattice gas systems. An example of that would be Rothmann and Keller's algorithm~\cite{rothman1994lattice} that is now known as the color gradient method. This method depends on neighboring site average density gradients.

The sampling collision operator takes inspiration from the BGK collision operator for the Boltzmann equation. The Boltzmann equation is an evolution equation for a distribution function $f_i(x,t)$, and the BGK collision operator is an approximation that simply moves the system towards it's local equilibrium distribution $f_i^0(x,t)$~\cite{BGK} where the local equilibrium distribution depends on the locally conserved quantities. This BGK lattice Boltzmann algorithm will be discussed in more detail in upcoming sections. 

In a lattice gas system, the system is described by $\{n_i\}_g$. Unlike in the lattice Boltzmann case, the local equilibrium state for the $n_i$ is not given by an equilibrium value, but rather by a probability distribution around some average value, known as the local equilibrium ensemble. This local equilibrium ensemble, given a local pre-collision set of occupation numbers $\{n_i\}_l$, gives a probability $P^0(\{\hat{n}_i\}_l; \{n_i\}_l)$ for each post-collision set $\{\hat{n}_i\}_l$. A post-collision set of occupation numbers can be then be sampled from the local equilibrium ensemble in a process discussed in detail in Seekins and Wagner's paper~\cite{seekins2022integer}. The collision operator for the sampling lattice gas algorithm thus becomes the difference between the pre-collision and post-collision occupation numbers, explicitly defined as,
\begin{equation}
    \Xi_i[\{n_j\}_l]=\hat{n}_i-n_i.
    \label{eqn: Sampling Col. Op.}
\end{equation}

However, this representation of the collision operator does not account for the fact that the system may not relax fully to local equilibrium in one timestep. To simulate this under-relaxation in a system of discrete particles, we can represent the fraction of particles that should undergo collision in that timestep as $\omega$. We create a subset of our occupation numbers $\{n^\omega_i\}_l$ to undergo collision, with the probability that any single particle is a member of the subset being given by $\omega$. Therefore, each specific value of $n_i^\omega$ has a probability associated with it given by the binomial distribution,
\begin{equation} 
    P(n^\omega_i) =\left( \begin{array}{c}
        n_i\\
        n^\omega_i
    \end{array} \right) \omega^{n^\omega_i} (1-\omega)^{n_i-n^\omega_i}.
\end{equation}

We are able to obtain the number of particles for each $n_i^\omega$ by sampling out of this binomial distribution. We create a pre-collision set $\{n^\omega_i\}_l$, and then use the probability distribution $P^0(\{\hat{n}^\omega_i\}_l; \{n^\omega_i\}_l)$ to sample a new, post collision set of occupation numbers $\{\hat{n}^\omega_i\}_l$. The collision operator for a specific value of $\omega$ is then given by,
\begin{equation}
    \Xi_i^\omega[\{n_j\}_l] = \hat{n}^\omega_i-n_i^\omega.
\end{equation}
All that remains to fully define the collision operator is to derive the local equilibrium ensemble explicitly.

\subsection{Derivation of the Local Equilibrium Ensemble}
In the hydrodynamic case considered here there are two conserved quantities: mass and momentum. Since momentum is conserved, the probability distribution for the $n_i$ will be correlated. However, by transforming our representation of the system from being expressed as an set of individual occupation numbers $\{n_i\}_l$ into being expressed by the hydrodynamic moments we may be able to obtain an uncorrelated probability distribution. These moments are physical quantities within the simulation, with a total number given by the total number of system velocities $Q$. From some fundamental particle collision process like that of Blommel and Wagner~\cite{blommel2018integer} we deduce the local equilibrium probability distributions for the non-conserved moments. We then use these local equilibrium ensembles to sample the post-collision moments as random variables. This is the extent of how far we are able to describe our algorithm in all generality. 

To know what moment representation is applicable to our system, we must focus our attention on a specific example. As this paper serves as a proof of concept for our sampling algorithm, we focus on the simplest possible system that recovers hydrodynamic effects, a one-dimensional system with three total velocity vectors, leaving the extension of the algorithm to more complex systems, including higher dimensional systems, to future work. 

Also known as a D1Q3 system, the velocity vectors of this system connect a given lattice point to its nearest neighbors, and include a zero velocity state. Numerically, we can represent this by restricting our velocities to the set $v_i\in\{-1, 0, 1\}$, with indices $i$ that are given by the value of the velocity they represent (\textit{i.e.} $v_{-1}=-1$). Knowing the structure of our system, we know that we have a total of three moments due to having a total of three velocities, and because two of our local system quantities are conserved, their corresponding moments must be conserved as well. Thus, we can define the local density $N$ and the local momentum $J$ as moments,
\begin{align}
    N&=n_{1}+n_0+n_{-1},
    \label{eqn: N D1Q3}\\
    J&=n_{1}-n_{-1}.
    \label{eqn: J D1Q3}
\end{align}
We define a third moment $\pi$ as,
\begin{equation}
    \pi=n_{1}+n_{-1},
    \label{eqn: Pi D1Q3}
\end{equation}
which, not being a conserved quantity, is the only moment that is allowed to change over the course of a local collision step. Each combination of these three moments fully describes a unique occupation state $\{n_i\}_l$, as Eqs. (\ref{eqn: N D1Q3}), (\ref{eqn: J D1Q3}), and (\ref{eqn: Pi D1Q3}) can be inverted to recover our system's occupation numbers,
\begin{align}
    n_1&=\frac{\pi+J}{2},
    \label{eqn: n1 Moment Def}\\
    n_0&=N-\pi,
    \label{eqn: n0 Moment Def}\\
    n_{-1}&=\frac{\pi-J}{2}.
    \label{eqn: nm1 Moment Def}
\end{align}

Thus, we can redefine $P^0(\{\hat{n}_i\}_l; \{n_i\}_l)$ in terms of $N$, $J$, and $\pi$ as $P^0(\hat{N},\hat{J},\hat{\pi}; N, J, \pi)$, or the probability of finding a system in a state with moments $\hat{N}$, $\hat{J}$, and $\hat{\pi}$ after a collision given a pre-collision state with moments $N$, $J$, and $\pi$. We can then simplify this representation, as our conservation rules state that $\hat{N}=N$ and $\hat{J}=J$. Furthermore, because we are assuming that the system is in local equilibrium when described by this distribution, we know that $\hat{\pi}$ does not depend on the pre-collision $\pi$ value. The probability distribution for our ensemble can then be written as $P^0(\pi; N, J)$, which is the equilibrium probability of finding the system in a post-collision state with a $\pi$ value of $\pi$, given a pre-collision system with conserved moments $N$ and $J$, and an arbitrary initial value for $\pi$.

We now derive the particular probability distribution that recovers the integer lattice gas results obtained by Blommel and Wagner. In local equilibrium, this distribution must be invariant under collisions. By using Blommel and Wagner's algorithm as a basis, we restrict the possible collisions of our system to only involving pairs of particles. These collisions conserve momentum, and their equilibrium ensemble must be stationary over each collision. This restricts us from altering states where two selected particles are moving in the same direction, such as $\{1,1\}$, where changing the velocity of the particles would violate momentum conservation in 1D systems. Since we are using momentum space to track the collision, any collisions that do not alter the $\pi$ value of the system have no effect, an example being $\{ 1,-1\}\to\{-1,1\}$. There are only four two-particle collisions for a D1Q3 system that alter the $\pi$ value: two that increase the $\pi$ value by two and two that decrease it by two,
\begin{equation}
    \pi \rightarrow \pi+2\biggr\{\begin{array}{c}
    \{0,0\}\to\{1,-1\}\\
    \{0,0\}\to\{-1,1\},\\
    \end{array} \hspace{0.2cm}
    \label{eqn: Pi Altering Col Up}
\end{equation}
\begin{equation}
    \pi+2 \rightarrow \pi\biggr\{\begin{array}{c}
    \{-1,1\}\to\{0,0\}\\
    \{1,-1\}\to\{0,0\}.\\
    \end{array} \hspace{0.3cm}
    \label{eqn: Pi Altering Col Down}
\end{equation}
The probabilities of each of these possible collisions occurring is given by the probability of picking a pair of particles with the proper set of velocity indices $\{i,j\}$, which is written as $p_{\{i,j\}}$, multiplied by the probability of the collision occurring from velocity state $\{i,j\}$ to velocity state $\{k,l\}$, written as $P_{\{i,j\}\to\{k,l\}}$. Thus, we can write $T_{\pi\to\pi+2}$, which is the probability of the system transitioning from a state with a $\pi$ value of $\pi$ to one with a $\pi$ value of $\pi+2$, as,
\begin{equation}
    T_{\pi\to\pi+2}=p_{\{0,0\}}P_{\{0,0\}\to \{1,-1\}}+p_{\{0,0\}}P_{\{0,0\}\to \{-1,1\}},
    \label{eqn: Tup}
\end{equation}
and $T_{\pi+2\to\pi}$, which is the probability of the system transitioning from a state with a $\pi$ value of $\pi+2$ to one with a $\pi$ value of $\pi$, as,
\begin{equation}
    T_{\pi+2\to\pi}=p_{\{1,-1\}}P_{\{1,-1\}\to \{0,0\}}+p_{\{-1,1\}}P_{\{-1,1\}\to \{0,0\}}.
    \label{eqn: Tdown}
\end{equation}

We assume that each particle has an equal likelihood to undergo collision. Thus, $p_{\{i,j\}}$ can be defined by considering the fraction of particles in the relevant $n_i$ occupation states compared to the total density at their lattice site $N$,
\begin{equation}
    p_{\{i,j\}}=\frac{n_i(n_j-\delta_{ij})}{N(N-1)},
\end{equation}
where $\delta_{ij}$ is the Kronecker delta, and is incorporated to avoid double counting particles should the indices $i$ and $j$ be the same. The $P_{\{i,j\}\to\{k,l\}}$ values were defined explicitly by Blommel and Wagner such that at zero velocity the typical lattice Boltzmann equilibrium values are recovered on average~\cite{blommel2018integer}. For the relevant collisions in Eqs. (\ref{eqn: Tup}) and (\ref{eqn: Tdown}), these values are given by,
\begin{align}
    &P_{\{0,0\}\to \{1,-1\}} = 1/16,\\
    &P_{\{0,0\} \to \{-1,1\}} = 1/16,\\
    &P_{\{1,-1\} \to \{0,0\}} = 1,\\
    &P_{\{-1,1\} \to \{0,0\}} = 1.
\end{align}
It is not obvious that the probability $P_{\{i,j\}\to \{k,l\}}$ should necessarily be independent of the locally conserved quantities. However, as Blommel and Wagner's method makes this assumption, we make it as well to adapt their algorithm. A true test of the validity of this approach could be done by looking at the coarse-graining of a molecular dynamics simulation, similar to the work done by Parsa \textit{et al.}~\cite{parsa2017lattice}.
 
We can now rewrite the transition probabilities as,
\begin{align}
   T_{\pi\to\pi+2}&=\frac{n_0(n_0-1)}{8N(N-1)},\\
   T_{\pi+2\to\pi}&=\frac{2n_1n_{-1}}{N(N-1)},
\end{align}
and transform them into moment representation by using Eqs. (\ref{eqn: n1 Moment Def}), (\ref{eqn: n0 Moment Def}), and (\ref{eqn: nm1 Moment Def}),
\begin{align}
    T_{\pi\to\pi+2}&=\frac{(N-\pi)(N-\pi-1)}{8N(N-1)},
    \label{eqn: TUpMoment}\\
    T_{\pi+2\to\pi}&=\frac{\left((\pi+2)^2-J^2\right)}{2N(N-1)}.
    \label{eqn: TDownMoment}
\end{align}

We can write,
\begin{equation}
    T_{\pi\to\pi+2}P^0(\pi; N, J)=T_{\pi+2\to\pi}P^0(\pi+2; N, J),
    \label{eqn: Detailed Balance}
\end{equation}
using the fact that $P^0(\pi; N, J)$ must be invariant under a single collision. Eq. (\ref{eqn: Detailed Balance}) can then be rewritten using Eqs. (\ref{eqn: TUpMoment}) and (\ref{eqn: TDownMoment}),
\begin{align}
    &\frac{(N-\pi)(N-\pi-1)}{8N(N-1)}P^0(\pi;N,J)=\\\nonumber
    &\frac{\left((\pi+2)^2-J^2\right)}{2N(N-1)}P^0(\pi+2;N,J).
    \label{eqn: Detailed Balance Final}
\end{align}

Simplifying this expression gives us a recursive formula for $P^0(\pi;N,J)$,
\begin{equation}
    P^0(\pi+2;N,J)=\frac{(N-\pi)(N-\pi-1)}{4((\pi+2)^2+J^2)}P^0(\pi;N,J).
    \label{eqn: Pi Dist Recursive}
\end{equation}
 This recursive formula, along with the requirement for normalization,
 \begin{equation}
     \sum_\pi P^0(\pi; N, J)=1,
 \end{equation}
 fully describes the local equilibrium ensemble. If we sample random $\hat{\pi}$ values out of this local equilibrium ensemble, our collision operator in moment space becomes,
 \begin{equation}
     \Xi_\pi=\hat{\pi}-\pi.
     \label{eqn: sampling col op pi}
 \end{equation}
 
 This representation of the collision operator is entirely equivalent to the representation in Eq. (\ref{eqn: Sampling Col. Op.}). To put this in the context of continuous fluctuating methods: the alteration of the $\pi$ value in the collision operator represents a microscopic implementation of the fluctuating stress tensor. Due to the required momentum conservation the local equilibrium ensemble will, starting at $\pi=J$, increment by two, and thus will maintain its initial parity. Initial $N$ and $J$ values of opposing parity result in a maximum possible $\pi$ value of $N-1$. Also, since the particles occupying an $n_0$ state do not stream between lattice sites, and are only altered in multiples of two, the parity of the $n_0$ value at each lattice site from initialization acts as a spurious conserved quantity, the (minor) consequences of which will be discussed in the upcoming sections.

The explicit formula corresponding to the recursive formula of our local equilibrium ensemble can be found by considering the evolution of the correct global equilibrium distribution, something that will be discussed at length in the next section, including a derivation of this explicit formula. This explicit formula is given by,
\begin{align}
    &\frac{1}{P^{0}(\pi; N, J)}=4^{N-\pi} \left(N-\pi\right)! \left(\frac{\pi+\left|J\right|}{2}\right)!\left(\frac{\pi-\left|J\right|}{2}\right)!\label{eqn: Local Explicit Formula}\\\nonumber
    &\sum_{k=0}^{\left\lfloor\frac{N-\left|J\right|}{2}\right\rfloor}\left(4^{N-2k-\left|J\right|} \left(N-2k-\left|J\right|\right)! \left(k+\left|J\right|\right)!\left(k\right)!\right)^{-1},
\end{align}
where $k$ is an index which enumerates the possible values of $\pi$ while taking into account the fact that those values increment by multiples of two, and the floor function $\lfloor...\rfloor$ returns the greatest integer value less than its argument.

\begin{figure} 
    \centering
    \includegraphics[width=\columnwidth, clip=true]{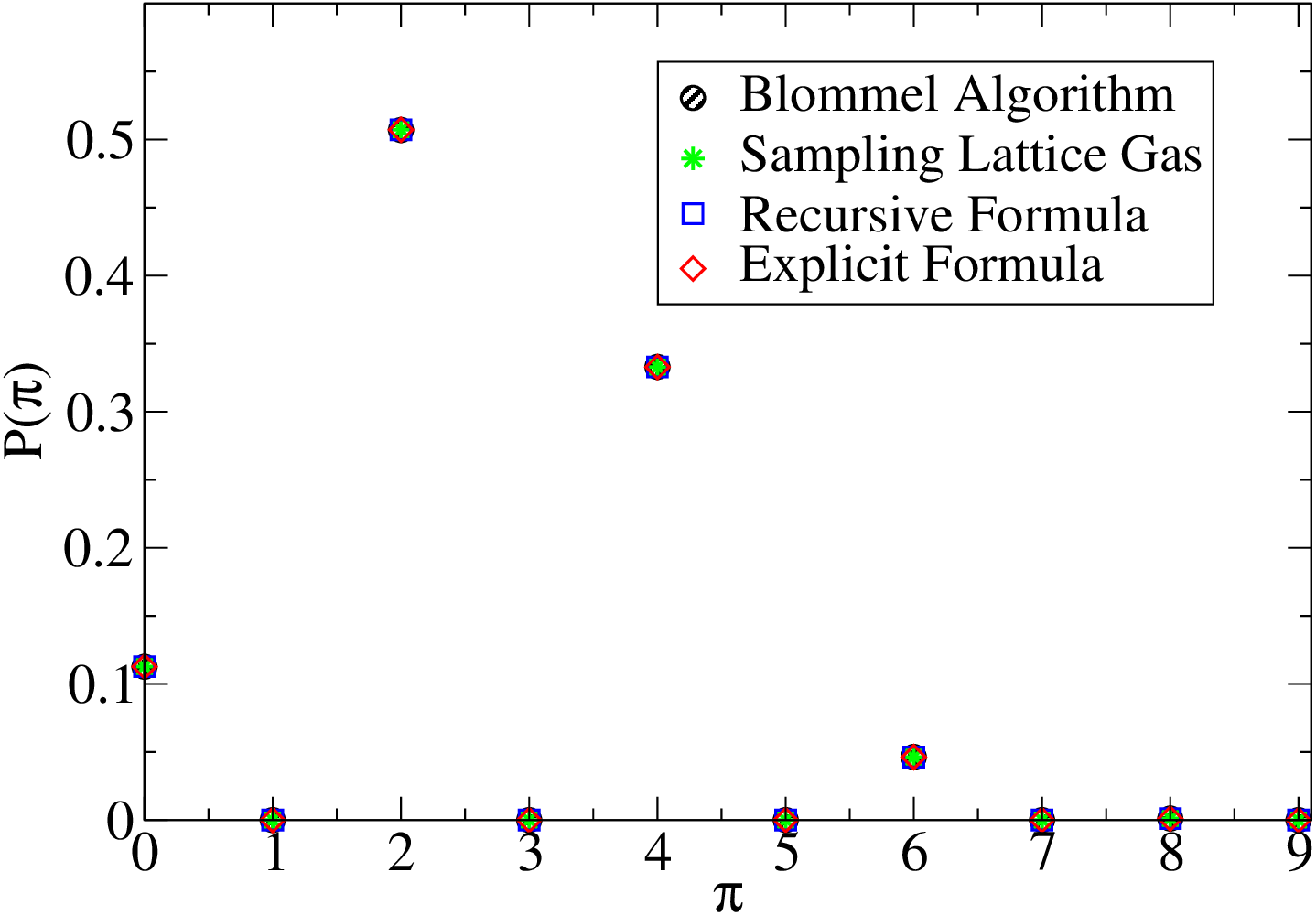}
    \caption{The Local equilbrium ensemble of $\pi$ values for $N=9$ and $J=0$. A single lattice site was made to undergo 100,000 collisions for both Blommel and Wagner's algorithm and our sampling algorithm ~\cite{blommel2018integer}. Then, a further 100,000,000 collisions were run, collecting the $\pi$ value after each. The histograms of these data sets are compared to the recursive and explicit formulae given by Eqs. (\ref{eqn: Pi Dist Recursive}) and (\ref{eqn: Local Explicit Formula}) respectively}.
    \label{fig: PiLocal}
\end{figure}
A comparison between this explicit formula and its recursive counterpart can be seen in Fig. \ref{fig: PiLocal}, along with histograms of local equilibrium $\pi$ values from both Blommel's algorithm and the sampling algorithm put forward in this paper. A low density system with a local $N$ value of $9$ was chosen due both to the fact that this local equilibrium ensemble is more distinct from a binomial distribution at lower densities and to the fact that we wished to display the lack of negative densities present in the distribution. 

The histograms of data from both Blommel and Wagner's algorithm and our sampling lattice gas algorithm show excellent agreement to the recursive and explicit formulae of the theoretical local equilibrium ensemble shown in Eqs. (\ref{eqn: Pi Dist Recursive}) and (\ref{eqn: Local Explicit Formula}), as well as maintaining the proper parity for their initial $J$ value. To ensure proper agreement, additional tests were run for different configurations of $N$ and $J$ values, finding results consistent with those shown in Fig. \ref{fig: PiLocal}. Thus, for brevity, we show only this single example.


\subsection{Global Equilibrium Ensemble Analysis and finding an Explicit Formula for the Local Equilibrium Ensemble\label{sec: GlobalDist}}
Analyzing the global equilibrium ensemble of our lattice gas system serves two primary purposes. Firstly, it allows us to better define our lattice gas algorithm's equilibrium state for the purpose of discussing the nature of its Boltzmann limit, which we will discuss in more detail in the next section. Secondly, it allows us to analytically derive an explicit formula for the local equilibrium ensemble, defined in the prior section as Eq. (\ref{eqn: Local Explicit Formula}).

Blommel and Wagner followed Adhikari \textit{et al.} \cite{adhikari2005fluctuating} in using arguments by Landau and Lifshitz~\cite[Sec. 114]{landau1969statistical} to postulate that the global equilibrium ensemble of a lattice gas should be given by a Poisson distribution around an equilibrium value~\cite{blommel2018integer}. Due to the spurious parity conservation of the $n_0$ value at each lattice site this probability distribution cannot be the correct equilibrium ensemble for any specific set of initial conditions. Instead, particular microstates have to be removed from the total Poisson distribution. However, Blommel and Wagner showed numerically that the occupation numbers recovered in equilibrium are approximately Poisson distributed when the initial conditions themselves give a random distribution of even and odd $n_0$ and the equilibrium is understood as an ensemble average over these initial conditions. 

The Poisson distribution of independent $n_i$ values around a given set of means $\tilde{f}_i$ is given by:
\begin{equation}
    P^{\mathrm{Pois}}(\{n_i\}_l, \tilde{f}_i)=\prod_i\exp\left(-\tilde{f}_i\right)\frac{(\tilde{f}_i)^{n_i}}{(n_i)!},
    \label{eqn: Poisson ni}
\end{equation}
where $P^{\mathrm{Pois}}(\{n_i\}_l, \tilde{f}_i)$ is the probability of finding a particular lattice site in state $\{n_i\}_l$. The $\tilde{f}_i$ values for Blommel and Wagner's algorithm are given by~\cite{blommel2018integer},
\begin{align}
    &f_i^{0}=
    \label{eqn: Entropic Local Value}\\
    &\bar{N} w_i\left(1+3v_iu+(3v_i^2-1)\left(\sqrt{1+3(u)^2}-1\right)\right),\nonumber
\end{align}
which is the equilibrium distribution for the entropic lattice Boltzmann method. The exact formulation of the ensemble average density $\bar{N}$ and velocity $u$ will be discussed in the next section, however in a global equilibrium system they are constant quantities that only depend on the initial conditions of that system,
\begin{align}
    \bar{N}^{eq}=\sum_i\sum_x \frac{n_i(x,0)}{L},
    \label{eqn: NbarEq}\\
    \bar{N}^{eq} u^{eq}=\bar{J}^{eq}=\sum_i\sum_x \frac{n_i(x,0)v_i}{L},
    \label{eqn: JbarEq}
\end{align}
where $L$ is the total lattice length of the system. This means that $f_i^{0}(\bar{N}^{eq},u^{eq})$ is a constant quantity.

We can write Eq. (\ref{eqn: Poisson ni}) in terms of the system moments by utilizing Eqs. (\ref{eqn: n1 Moment Def}), (\ref{eqn: n0 Moment Def}), and (\ref{eqn: nm1 Moment Def}). For brevity, the $n_i$ values are kept as $n_i(N, J, \pi)$ until otherwise required for calculation,
\begin{align}
    &P^{\mathrm{Pois}}[\{N,J,\pi\}_l, f_i^{eq}(\bar{N}^{eq}, \bar{J}^{eq})] = \\
    &\prod_i \exp[-f_i^{eq}(\bar{N}^{eq},\bar{J}^{eq})]\frac{f_i^{eq}(\bar{N}^{eq}, \bar{J}^{eq})^{n_i(N,J,\pi)}}{n_i(N,J,\pi)!}.
    \label{eqn: Peq}\nonumber
\end{align}
Here $u^{eq}$ is replaced by $\bar{J}^{eq}$, as $u^{eq}$ can be written as $\bar{J}^{eq}/\bar{N}^{eq}$.

We can further analyze this Poisson distribution by considering the requirements for equilibrium. Specifically, we require that our global ensemble must be invariant under Eq. (\ref{eqn: lgEvo}). As the $n_i$ values from Eq. (\ref{eqn: Poisson ni}) are independently distributed, this distribution is inherently invariant under streaming. Therefore, it only remains to be proven that the system is invariant under collisions. We can test this analytically by considering the effect of a single collision on the probability distribution $P^{\mathrm{Pois}}[\{N,J,\pi\}_l,f_i^{eq}(\bar{N}^{eq}, \bar{J}^{eq})]$ as we did in Eq. (\ref{eqn: Detailed Balance}), while adjusting for the fact that the effect of the collision operator is only required to be zero on average. Thus, we can write,
\begin{align}
    &\sum_{N,J}P^{\mathrm{Pois}}[\{N,J,\pi\}_l,f_i^{eq}(\bar{N}^{eq},\bar{J}^{eq})]T_{\pi\to\pi+2}=\\\nonumber
    &\sum_{N,J}P^{\mathrm{Pois}}[\{N,J,\pi+2\}_l,f(\bar{N}^{eq},\bar{J}^{eq})T_{\pi+2\to\pi},
\end{align}
where the sum $\sum_{N,J}$ is a sum over all possible values of $N$ and $J$, with minimum values $N=\pi$ and $J=-\pi$ and maxima of the number of total particles in the system for $N$, and $J=\pi$. Substituting in our previously defined quantities on both sides of this equation gives us,
\begin{align}
    &\sum_{N,J}\prod_i \exp(-f_i^{eq})\frac{(f_i^{eq})^{n_i(N,J,\pi)}}{n_i(N,J,\pi)!}\frac{(N-\pi)(N-\pi-1)}{8N(N-1)}=
    \label{eqn: Global DB Mid}\\
    &\sum_{N,J}\prod_i \exp(-f_i^{eq})\frac{(f_i^{eq})^{n_i(N,J,\pi+2)}}{n_i(N,J,\pi+2)!}\frac{(\pi^2-J^2)}{2N(N-1)}.\nonumber   
\end{align}

We can further simplify this by taking the product over $i$, and substituting in Eqs. (\ref{eqn: n1 Moment Def}), (\ref{eqn: n0 Moment Def}), and (\ref{eqn: nm1 Moment Def}) for the remaining $n_i$ values. However, for the sake of clarity, we first define factors $\alpha$ and $\beta$ as common terms on both sides of the equation after the substitution,
\begin{align}
    \alpha&=\frac{1}{4}\exp(-f_0^{eq}-f_1^{eq}-f_{-1}^{eq})\frac{1}{N(N-1)},\\
    \beta&=\frac{(f_0^{eq})^{N-\pi}(f_1^{eq})^{\frac{\pi+J}{2}}(f_{-1}^{eq})^{\frac{\pi-J}{2}}}{\left(N-\pi\right)!\left(\frac{\pi+J}{2}\right)!\left(\frac{\pi-J}{2}\right)!}.
\end{align}
Thus, Eq. (\ref{eqn: Global DB Mid}) becomes,
\begin{align}
    &\sum_{N,J}\frac{1}{2}\alpha\beta\left(N-\pi\right)\left(N-\pi-1\right)= 
    \label{eqn: Global DB Late}\\
    &\sum_{N,J}8\alpha\beta\frac{f^{eq}_1f^{eq}_{-1}}{(f^{eq}_0)^2}\left(N-\pi\right)\left(N-\pi-1\right).\nonumber
\end{align}

This equation can be simplified once more by calculating $f_1^{eq}f_{-1}^{eq}/(f_0^{eq})^2$, which can be calculated using Eq. (\ref{eqn: Entropic Local Value}), and comes out to a value of $1/16$. This, when substituted into Eq. (\ref{eqn: Global DB Late}), makes both sides of the equation identical, which proves that $P^{\mathrm{Pois}}[\{N,J,\pi\}_l,f_i^{eq}(\bar{N}^{eq}, \bar{J}^{eq})]$ is invariant under collisions. In fact, because we proved equivalence for each individual term of the sum over $N$ and $J$, the aforementioned Poisson distribution must be a representation of the local equilibrium ensemble for our lattice gas when considered for fixed $N$ and $J$,. Specifically, we find,
\begin{equation}
    P^0(\pi;N,J)=\gamma(\pi)\frac{1}{\sum^N_{\pi'=|J|}\gamma(\pi')}.
    \label{eqn: P0GammaNorm}
\end{equation}
where $\gamma(\pi)$ is given by
\begin{equation}
    \gamma(\pi)=\mathrm{exp}\left(f_0^{eq}f_1^{eq}f_{-1}^{eq}\right)(\frac{f_0^{eq})^{N-\pi}(f_1^{eq})^{\frac{\pi+J}{2}}(f_{-1}^{eq})^{\frac{\pi-J}{2}}}{(N-\pi)!\left(\frac{\pi+J}{2}\right)!\left(\frac{\pi-J}{2}\right)!}.
    \label{eqn: Gamma}
\end{equation}

Because Eq. (\ref{eqn: Gamma}) does not include all possible microstates of the Poisson distribution, it is no longer normalized. To account for this, Eq. (\ref{eqn: P0GammaNorm}) includes a normalization factor. This normalization factor consists of a sum over all possible values of $\pi$, utilizing an index $\pi'$ to do so. However this representation does not take into account that the $\pi$ value can only be incremented by multiples of two. By defining a second index $k$, and setting $\pi'=2k+|J|$, we can find a system that increases by two between values of $\pi'$ with a lower bound of $0$, and an upper bound of $\lfloor(N-|J|)/2\rfloor$. The floor function takes into account the fact that the maximum value of $\pi$ depends on the relative parity of $N$ and $J$. 

Blommel and Wagner make the assumption when defining their collision rules that only local quantities can affect the local collisions~\cite{blommel2018integer}. Thus, as a representation of the local equilibrium ensemble of Blommel and Wagner's algorithm, Eq. (\ref{eqn: P0GammaNorm}) must only depend on local quantities. Thus, we must be able to rewrite Eq. (\ref{eqn: P0GammaNorm}) such that the dependency on the $f_i^{eq}$ values, which depend on the global quantities $\bar{N}^{eq}$ and $\bar{J}^{eq}$, vanishes. By using the aforementioned relation that $f_1^{eq}f_{-1}^{eq}/(f_0^{eq})^2=1/16$ to substitute $f_{0}=\sqrt{16f_1^{eq}f_{-1}^{eq}}$, we can rewrite $\gamma(\pi)$ as,
\begin{align}
    &\gamma(\pi)=\\
    &\mathrm{exp}\left[4(f_1^{eq}f_{-1}^{eq})^{3/2}\right]\frac{(f_1^{eq})^{\frac{N+J}{2}}(f_{-1}^{eq})^{\frac{N-J}{2}}}{4^{N-\pi}(N-\pi)!\left(\frac{\pi+J}{2}\right)!\left(\frac{\pi-J}{2}\right)!}.\nonumber
\end{align}
In this form, $\gamma(\pi)$ has all $f_i^{eq}$ terms independent of $\pi$, meaning that when the whole of Eq. (\ref{eqn: P0GammaNorm}) considered, the $f_i^{eq}$ terms will cancel out due to the normalization factor, which, along with making the aforementioned index substitution for $\pi'$, recovers Eq. (\ref{eqn: Local Explicit Formula}).

A collection of these local equilibrium ensembles can then in theory form the proper global equilibrium ensemble for our system, taking into account the spuriously conserved $n_0$ parity by picking only $N$ and $J$ values for the global equilibrium ensemble that maintain this conservation.


\section{The Boltzmann Limit and its Uniqueness\label{sec: Boltzmann Algorithm}}
To derive the hydrodynamic limit of our lattice gas algorithm, which will be given by a set of continuity and Navier-Stokes equations, it is required that we first know the Boltzmann limit of our lattice gas. The task of obtaining a Boltzmann limit for a lattice gas goes back to the very origin of lattice gases \cite{frisch1987lattice}. Even in the continuous case of obtaining the Boltzmann equation from the BBGKY hierarchy \cite{Huang1987Statistical} it is necessary to make the significant assumption that correlations can be neglected, and that the two particle distribution function can be factorized as the product of two one-particle distribution functions. Usually it is assumed that it is the pre-collision two-particle distribution function that is uncorrelated, and this approximation introduces the irreversibility absent in the full BBGKY hierarchy. It is understood that this approximation will only be approximately valid, and the same difficulty exists for the Boltzmann limit of our lattice gas, as we will see below.

To define the Boltzmann limit of a lattice gas, we consider the ensemble average over the possible microstates of a lattice gas system given a specific initial state and elapsed time. The above mentioned difficulty is then moved to the task of defining the appropriate ensemble over which to average. For the Boolean lattice gases the expectation value $\langle n_i\rangle$ already defines the distribution of the $n_i$ completely and it only remains to make the assumption $\langle n_i n_j\rangle = \langle n_i\rangle \langle n_j\rangle$ \cite{frisch1987lattice}. In the case of an integer lattice gas just knowing $\langle n_i\rangle$ does not imply the underlying distribution function, giving more freedom in the definition of the ensemble over which to average.

The Boltzmann limit of the lattice gas evolution equation then gives us the corresponding lattice Boltzmann equation. A Chapman-Enskog analysis (or just moment equation~\cite{wagnerthesis}) of the lattice Boltzmann evolution equation~\cite{chapman1990mathematical} then recovers the continuity and Navier-Stokes equations of the corresponding lattice Boltzmann model.

For a lattice gas defined by the evolution equation given by Eq. (\ref{eqn: lgEvo}), we can formally find its Boltzmann limit by taking the ensemble average over this evolution equation. We start by defining,
\begin{equation}
    f_i(x, t)=\langle n_i(x,t)\rangle,
    \label{eqn: fi def}
\end{equation}
where the non-equilibrium ensemble average $\langle...\rangle$ is defined as,
\begin{equation}
    \langle X\rangle=\sum_{\{n_j\}_g} X P(\{n_j\}_g,t),
    \label{eqn: Boltzmann Avg}
\end{equation}
with $P(\{n_j\}_g,t)$ being the probability of finding the system in the specific microstate $\{n_j\}_g$ at time $t$. We can then take the Boltzmann average of Eq. (\ref{eqn: lgEvo}),
\begin{equation}
    \langle n_i(x+v_i\Delta t,t+\Delta t)\rangle = \langle n_i(x, t)\rangle+\langle\Xi_i(\{n_j\}_g)\rangle,
    \label{eqn: B Average ni}
\end{equation}
which  gives,
\begin{equation}
    f_i(x+v_i\Delta t, t+\Delta t) = f_i(x,t)+\Omega_i,
    \label{eqn: lBEvo}
\end{equation}
which is the lattice Boltzmann evolution equation associated with our lattice gas system. $\Omega_i$ in this case is our lattice Boltzmann collision operator, which is defined as the Boltzmann average of $\Xi_i$,
\begin{equation}
    \Omega_i=\langle\Xi_i(\{n_j\}_g)\rangle.
\end{equation}

We can also derive the $f_i$ values for any system for which we know the ensemble average $N$, $J$, and $\pi$ values mentioned in the prior section,
\begin{align}
    &\bar{N}(x,t)=\langle N \rangle,\label{eqn: BarN}\\
    &\bar{N}u=\bar{J}(x,t)=\langle J\rangle,\label{eqn: BarJ}\\
    &\bar{\pi}(x,t)=\langle \pi \rangle.\label{eqn: barPi}
\end{align}
This is possible by taking the Boltzmann average in Eq. (\ref{eqn: Boltzmann Avg}) over Eqs. (\ref{eqn: n1 Moment Def}), (\ref{eqn: n0 Moment Def}), and (\ref{eqn: nm1 Moment Def}). These expressions can then be written as,
\begin{align}
    &f_{1}=\frac{\bar{\pi}+\bar{J}}{2},\label{eqn: f1FromAvg}\\
    &f_{0}=\bar{N}-\bar{\pi},\label{eqn: f0FromAvg}\\
    &f_{-1}=\frac{\bar{\pi}-\bar{J}}{2}.\label{eqn: fm1FromAvg}
\end{align}

For our algorithm to have a unique Boltzmann limit, we must be able to derive an expression for $\Omega_i$ that depends only on the local $f_i$ values. The difficulty that arises is that the $f_i$ do not uniquely define the $P(\{n_j\}_g,t)$. The molecular chaos assumption usually applied can be written as
  \begin{equation}
    P(\{n_j\}_g,t)\approx\prod_x \prod_i P(n_i,x,t)
    \label{eqn:factor}
  \end{equation}
  where the $P(n_i,x,t)$ is the probability for the occupation number $n_i(x,t)$ to occur.

Now for the Boolean lattice gas knowing $f_i$ is sufficient to know the one particle distribution function, which is simply given by $P(n_i\equiv 1,x,t)=f_i$ and $P(n_i\equiv 0,x,t)=1-f_i(x,t)$~\cite{frisch1987lattice, Broadwell2007Shock}. When velocity states are allowed to have more than a single particle occupying them, the $f_i$ values no longer fully define the $n_i$ distribution, even for uncorrelated particle distributions, and we will consider the effect of this freedom below.

Given the general discussion above, it  may seem surprising that a unique Boltzmann limit was found for the diffusive sampling lattice gas both by Chopard and Droz~\cite[sec. 5.7]{chopardcellular} and by Seekins and Wagner~\cite{seekins2022integer}. This case is subtly simpler, because collisions can be written as one-particle collisions with an imagined substrate, making correlations irrelevant in the derivation of the collision operator. Since all particles equally contribute to collisions, independent of their current velocity state, the form of the one-particle distribution is likewise irrelevant. Both groups found that the diffusive sampling lattice gas recovered the BGK lattice Boltzmann collision operator in their Boltzmann limit, given by,
\begin{equation}
    \Omega_i^{\mathrm{BGK}}=-\omega_{LB}\left(f_i-f_i^{0}\right).
    \label{eqn: BGK}
\end{equation}
This lattice Boltzmann collision operator sets the particle distribution $\{f_i\}_l$ to some local equilibrium distribution $\{f_i^{0}\}_l$, at a rate given by the relaxation time, given by the inverse of $\omega_{LB}$, which is analogous to the particle collision fraction $\omega$ in a lattice gas. In the Boltzmann limit of the diffusive sampling lattice gas, $f_i^0$ is the polynomial lattice Boltzmann equilibrium distribution.

The full analytical derivation of the collision operator in the hydrodynamic case is an interesting problem. While we looked at this problem, and some progress appears to be possible, it is outside the scope of the current paper. Instead we will numerically evaluate the effect of different one particle distribution functions corresponding to the same set of expectation values $f_i$ on the collision operator, and we will see that this choice can indeed affect the collision operator. We also show that more significant effects are possible if the factorization assumption of Eq. (\ref{eqn:factor}) is dropped. This is important to make sense of some of  our simulations shown below that resulted in a significant deviation from the BGK collision operator.

Blommel and Wagner showed that the BGK collision operator using the entropic local equilibrium distribution given by Eq. (\ref{eqn: Entropic Local Value}) is a good approximation in most cases~\cite{blommel2018integer}. To test this for our sampling algorithm, we can compare the average collision operators for different ensembles by looking at the difference in the average pre- and post-collision $\pi$ values that result from applying the collision operator to those initial distributions.

The first ensemble we consider is actually a fixed value for the $n_i$, given by $n_i=f_i$, when the $f_i$ happen to be integers and if they aren't integers then we interpolate them such that the mean values recover the non-integer $f_i$ by defining
\begin{align}
    &P^{\delta}(\{n_i\}_l)=\label{eqn: P Delta}\\
    &\prod_i(f_i-\lfloor f_i\rfloor)\delta(n_i-\lfloor f_i\rfloor+1)+(1-(f_i-\lfloor f_i\rfloor))\delta(n_i-\lfloor f_i\rfloor).\label{eqn:delta}
\end{align}
 This is what we refer to as the ``Delta distributed $n_i$'' below. The second distribution we consider is a Poisson distributed set of $n_i$ which defines the probability of such states as
  \begin{equation}
    P^{Poisson}(\{n_j\})=\prod_i e^{-f_i}\frac{n_i^{f_i}}{n_i!}.
    \label{eqn:poisson}
  \end{equation}
  This particular distribution would recover an equilibrium distribution when $\langle \pi\rangle$ corresponds to the equilibrium value.
  Each of these distributions fulfills our requirement that $\langle n_i\rangle = f_i$. We will see below that collisions applied to these ensembles are close to, but not identical to, the BGK collision operator.

In some of the simulations presented below we observed deviations that we were too large to be explained by different uncorreleated distributions, like those of Eqn. (\ref{eqn:delta}) and Eqn. (\ref{eqn:poisson}). We therefore also wanted to consider correlated distributions.  Specifically, we considered alternative sets of delta distributed particles with two highly correlated states. In both cases, either all particles at a given lattice site are moving, or all particles at that lattice site are at rest. The probability for a given lattice site to be in a moving state is given by the ratio of $\bar{\pi}$ to $\bar{N}$. For the first probability distribution, the moving particles are evenly distributed within a single lattice site, after taking the initial $\bar{J}$ value into account, giving us,
\begin{align}
    &P_{\mathrm{corr 1}}(\{n_i\}_l; \bar{N}, \bar{J}, \bar{\pi})=
   \nonumber \\
    &\frac{\bar{\pi}}{\bar{N}}\delta\left(n_1-\frac{\bar{N}+\bar{J}}{2}\right)\delta\left(n_{-1}-\frac{\bar{N}-\bar{J}}{2}\right)\delta(n_0)\nonumber\\
    &+\left(1-\frac{\bar{\pi}}{\bar{N}}\right)\delta(n_1)\delta\left(n_{-1}\right)\delta\left(n_0-\bar{N}\right).\label{eqn: CorrP1}
\end{align}

For the second probability distribution, all lattice sites either have only left moving particles, only right moving particles, or only rest particles, with each lattice site chosen with moving particles having a probability of $0.5+\bar{J}/\bar{N}$ of having only rightward moving particles. Putting this together gives us,
\begin{align}
    &P_{\mathrm{corr2}}(\{n_i\}_l; \bar{N}, \bar{J}, \bar{\pi})=
   \nonumber\\ 
   &\frac{\bar{\pi}}{\bar{N}}\Biggl(\left[0.5+\frac{\bar{J}}{\bar{N}}\right]\delta\left[n_1-\bar{N}\right]\delta\left[n_{0}\right]\delta\left[n_{-1}\right]+\nonumber\\
   &\left[0.5-\frac{\bar{J}}{\bar{N}}\right]\delta\left[n_{-1}-\bar{N}\right]\delta\left[n_{0}\right]\delta\left[n_{1}\right]\Biggr)\nonumber\\
    &+\left(1-\frac{\bar{\pi}}{\bar{N}}\right)\delta\left(n_0-\bar{N}\right)\delta\left(n_{1}\right)\delta\left(n_{-1}\right). 
    \label{eqn: CorrP2}
\end{align}
As collisions require either a pair of rest particles or a pair of moving particles with opposing velocities, we can expect the number of collisions present in a system with a distribution given by $P_{\mathrm{corr2}}$ to be heavily diminished compared to other distributions, as well as only allowing for an increase in the total $\pi$ value of the system.

To test the Boltzmann limit of our collision operator, we initialized four systems of 100,000 lattice sites utilizing the Poisson distribution given by Eq. (\ref{eqn: Poisson ni}), the set of delta functions given by Eq. (\ref{eqn: P Delta}), and the two highly correlated distributions $P_\mathrm{corr1}$ and $P_\mathrm{corr2}$. We started by defining constant (independent of $x$) initial $\bar{N}$ and $\bar{J}$ values and setting the initial $\bar{\pi}$ value to its minimum ($\bar{\pi}=0$). We were then able to define $f_i$ values corresponding to those average quantities using Eqs. (\ref{eqn: f1FromAvg}), (\ref{eqn: f0FromAvg}), and (\ref{eqn: fm1FromAvg}). Each system was then collided without streaming, collecting each pre- and post-collision set of $\pi$ values, and the average of the resulting difference between the two was taken to be our averaged collision operator. 

To compare with the lattice Boltzmann, the BGK collision operator was calculated directly utilizing the $\{f_i\}_l$ in Eq. (\ref{eqn: BGK}) and with an equilibrium distribution given by Eq. (\ref{eqn: Entropic Local Value}). All simulations were then repeated, increasing the initial $\bar{\pi}$ value by 1 until the $\bar{\pi}$ value reached the $\bar{N}$ value selected.

\begin{figure} 
    \centering
    \includegraphics[width=\columnwidth,clip=true]{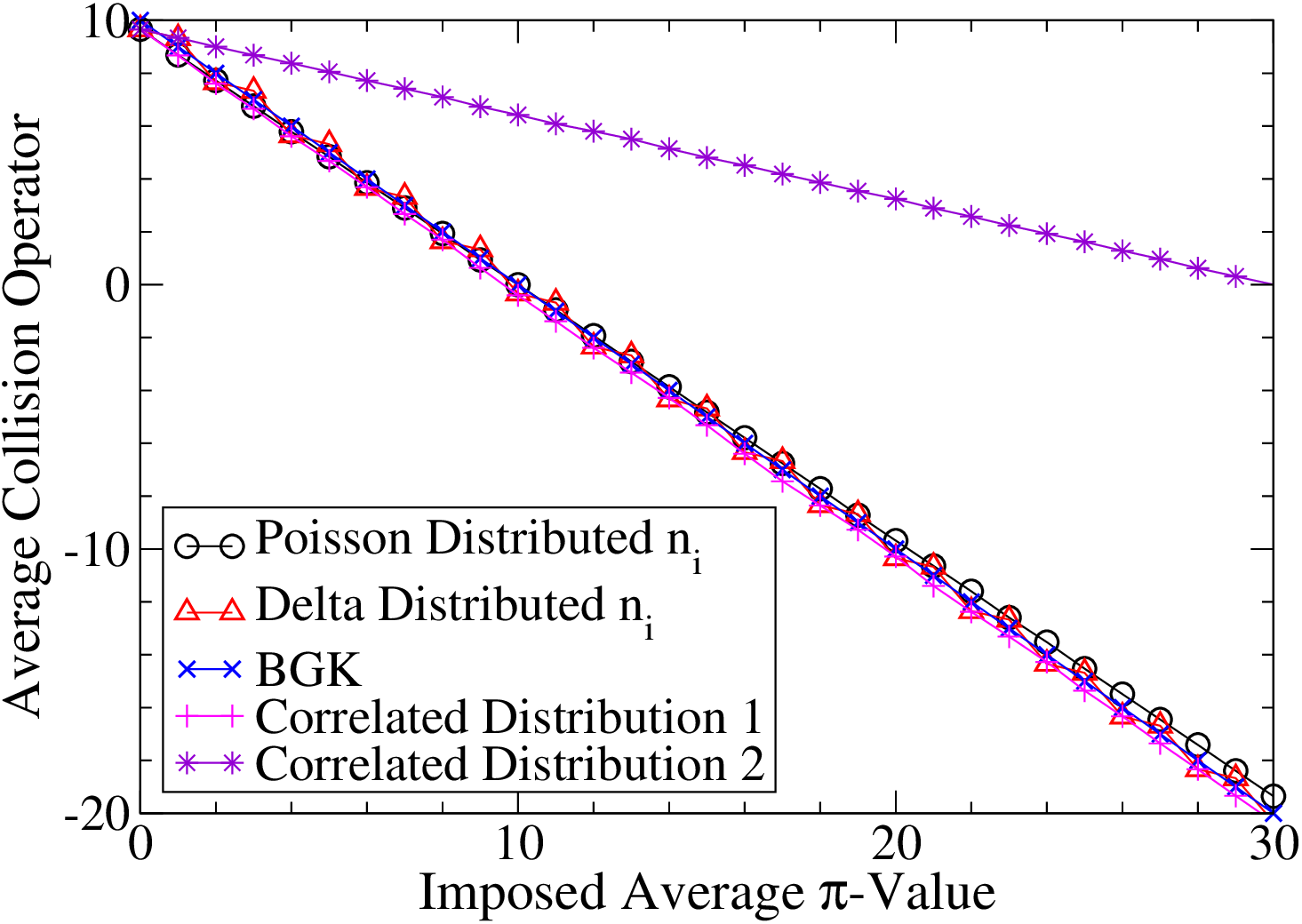}\\(a)\\
    \includegraphics[width=\columnwidth,clip=true]{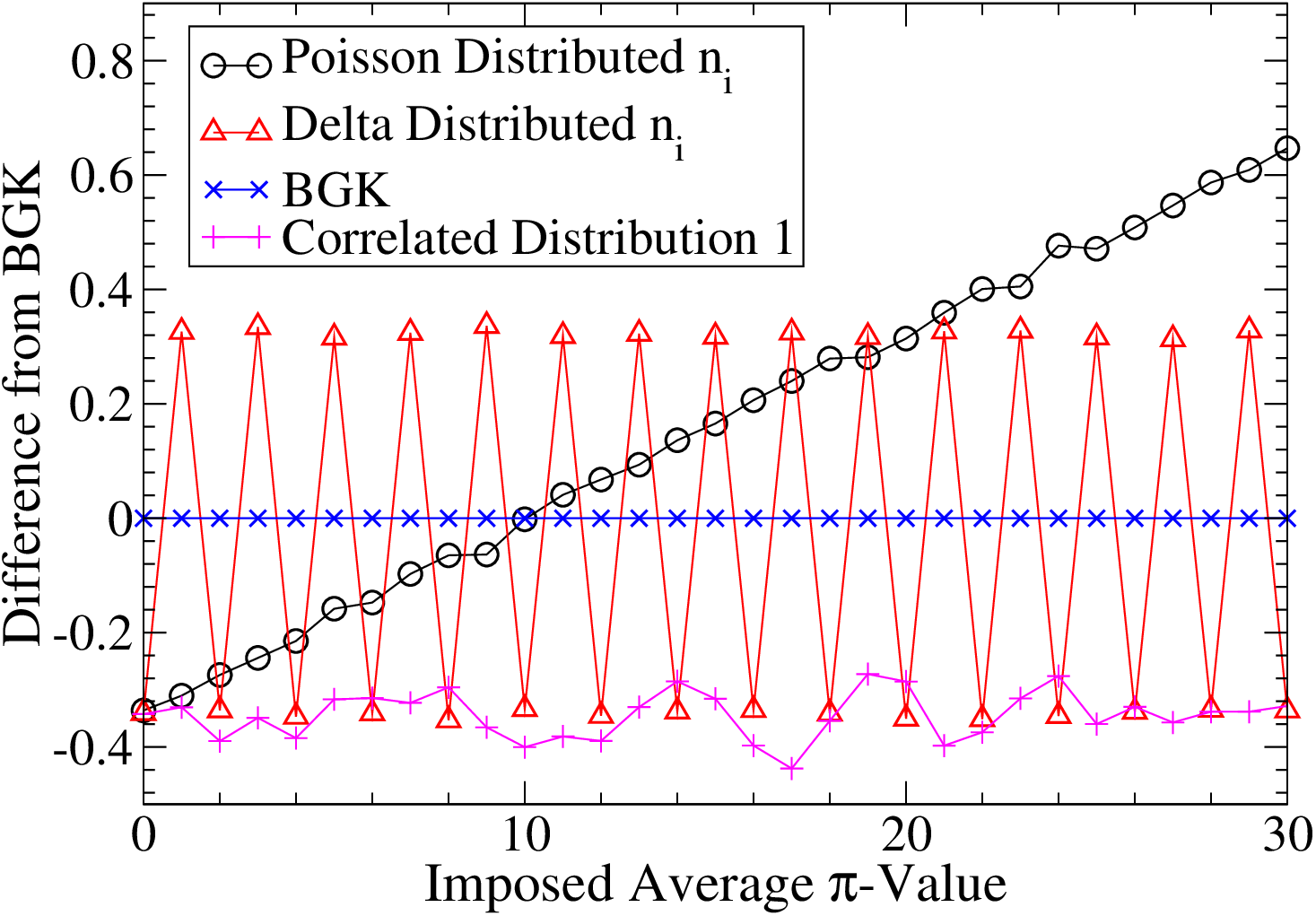}\\(b)
    \caption{The averaged collision operator given an imposed average $\pi$ value for four lattice gas particle distributions compared to the BGK collision operator given an initial average ensemble of $\{\bar{N}=30,\bar{J}=0,\bar{\pi}=0\}$ with $\omega=\omega_{LB}=1$}. Both the average collision operator (a) and the difference from the BGK calculated by subtracting the BGK from each collision operator (b) are shown. See text for further details.
    \label{fig: ColOpLinesUnscaled}
\end{figure}

Fig. \ref{fig: ColOpLinesUnscaled} shows the results of this test for a series of increasing average $\pi$ values. Fig. \ref{fig: ColOpLinesUnscaled} (a) shows the average change in $\pi$ value over the course of a single collision, and Fig. \ref{fig: ColOpLinesUnscaled} (b) shows the difference between each collision operator and the BGK collision operator.

The highly correlated initial condition of Eq. (\ref{eqn: CorrP2}) shows a behavior that is notably non-BGK. The equilibrium value of $\pi$ doesn't correspond to a zero crossing for the collision operator, meaning that $\pi$ -alues already larger than the equilibrium value will initially grow for this distribution. Clearly this move away from equilibrium will change the distribution so that the $\pi$ value will eventually decrease towards the equilibrium value. However, it is quite unclear how such non-BGK behavior will affect the hydrodynamic transport parameters.

Though the three uncorrelated collision operators and the collision operator resultant from $P_{\mathrm{corr1}}$ are similar, as can be seen in Fig. \ref{fig: ColOpLinesUnscaled} (a), they are not identical, which Fig. \ref{fig: ColOpLinesUnscaled} (b) shows more explicitly. We observe an oscillatory behavior of the delta distributed system. This is related to a slightly different average $\pi$ value for even or odd $n_0$. The slightly altered slope of the Poisson distributed system from the BGK and delta distributed systems can be interpreted as a slightly altered BGK relaxation value for $\omega_{LB}$. Importantly here the equilibrium value for $\pi$ is unaffected.

These results show that the Boltzmann limit of our algorithm is not unique. However, for the uncorrelated systems studied, the BGK entropic lattice Boltzmann algorithm is a good approximation. This entropic lattice Boltzmann algorithm is identical to the algorithm first developed by Ansumali \textit{et al.} \cite{Ansumali_2003}.

Using the previously discussed Chapman-Enskog analysis, the approximate continuity and Navier-Stokes equations for the D1Q3 sampling lattice gas can be derived as~\cite{Ansumali_2003},
\begin{equation}
    \partial_t\bar{N}+\partial_x\bar{J}=0,
    \label{eqn: continuity Eq}
\end{equation}
and,
\begin{equation}
    \partial_t\bar{J}+\partial_x\frac{\bar{J}^2}{\bar{N}}=\partial_x\bar{N}\theta+\frac{4}{3}\partial_x\nu_\omega\bar{N}\partial_x\frac{\bar{J}}{\bar{N}}.
    \label{eqn: NS Eq}
\end{equation}
Here $\nu_\omega$ is the kinematic viscosity related to the system's single-timestep particle collision fraction,
\begin{equation}
    \nu_\omega=\theta\left(\frac{1}{\omega_{LB}}-\frac{1}{2}\right),
    \label{eqn: Kinematic Viscosity}
\end{equation}
where $\theta$ is related to the speed of sound on our lattice,
\begin{equation}
    \theta= c_s^2=\sum_i f_i^{eq}(v_i-u)^2/\sum_i f_i^{eq}\approx\frac{1}{3}.
    \label{eqn: ThetaSoundSpeed}
\end{equation}
Blommel and Wagner show in Fig. 5 of their paper that this value for $\theta$ is only a good approximation for $|u|<0.4$~\cite{blommel2018integer}. Outside of that range, $\theta$ will have some dependence on the density, and thus the Navier-Stokes equation will have pressure and viscosity terms that are density dependent.

The kinematic viscosity in Eq. (\ref{eqn: Kinematic Viscosity}) can be lowered by selecting an $\omega_{LB}$ value between $1$ and $2$, which is useful for simulating low-viscosity systems. This is easily done for the BGK collision operator of Eq. (\ref{eqn: BGK}). These systems are generally referred to as overrelaxed. However for the lattice gas the collision probability $\omega$ has to be between zero and one. Some work has been done on this problem by Strand and Wagner~\cite{strand2022overrelaxation}, who created a method to simulate overrelaxation in diffusive systems, but their approach cannot be extended to hydrodynamic systems. Instead an additional step in the lattice gas approach has to be implemented, and even then the results cannot access the zero viscosity limit of $\omega=2$ for a finite number of particles per lattice cell. This extension, however, is outside the scope of this paper and will be addressed in a forthcoming publication.

\section{Non-equilibrium Dynamics}
In this section we will apply our algorithm to two specific cases to see when the BGK approximation for the lattice Boltzmann limit of our lattice gas is an appropriate approximation, and in what cases we see divergent behavior. We will find that over-all the agreement is very good, but there clearly are some cases where significant deviations are observed. We then examine a strongly non-equilibrium shock wave and find surprisingly close agreement between the BGK LBM approximation and the Boltzmann limit of the lattice gas.

\subsection{The Decaying Sound Wave}
To test the approximate Boltzmann limit of our lattice gas, we start by analyzing the decaying isothermal sound wave. While the isothermal nature makes this a non-physical application, it is still valuable to compare the lattice Boltzmann and lattice gas algorithms without addressing the complication of including energy conservation. An analytical solution for this system was found by Luiz Czelusniak~\cite{luizThesis} in his thesis, and is given by,
\begin{align}
    \bar{N}_{\mathrm{sound}}(x,t)&=\bar{N}^{eq}+A_0\sin\left(\frac{2\tilde{\pi}x}{L}\right)\cos(ft)e^{-\lambda t},
    \label{eqn: SineAnalyticalN}\\
    u_{\mathrm{sound}}&=B\cos\left(\frac{2\tilde{\pi}x}{L}\right)\sin\left(ft\right)e^{-\lambda t}.
    \label{eqn: SineAnalyticalu}
\end{align}
where $A_0$ is the initial amplitude of the wave. $\lambda$ is the exponential decay rate, which depends on the kinematic viscosity of the system such that,
\begin{equation}
    \lambda=\left(\frac{2\tilde{\pi}}{L}\right)^2\nu
    \label{eqn: nuAndlambda}
\end{equation}
$f$ is the oscillation frequency of the sine wave given by,
\begin{equation}
    f^2=\left(\frac{2\tilde{\pi}}{L}\right)^2\theta-\lambda^2,
\end{equation}
and $B$ is a constant that depends on the system's initial conditions,
\begin{equation}
    B=\frac{-A_0\sqrt{\theta}}{\bar{N}^{eq}}.
\end{equation}
Due to the fact that $\pi$ is already defined as a moment in our system, $\tilde{\pi}$ represents the mathematical constant. A key assumption for this analytical solution is that $J$ is small, so that the non-linear term $\partial_x\bar{J}^2/\bar{N}$ can be dropped in Eq. (\ref{eqn: NS Eq}). Thus, we are restricted to small values for $A_0$, on the order of $1\%$ of $\bar{N}^{eq}$. 

The sound wave must be initialized utilizing a particle distribution for lattice gas systems due to the discrete nature of particles. As the global equilibrium ensemble for the lattice gas is approximately Poisson distributed, we initialize our system by sampling the occupation numbers out of the Poisson distribution from Eq. (\ref{eqn: Poisson ni}), with $\tilde{f_i}$ values given by Eq. (\ref{eqn: Entropic Local Value}). The $\bar{N}$ and $u$ values needed to find the $\tilde{f_i}$ values can then be defined as the initial conditions for Eqs. (\ref{eqn: SineAnalyticalN}) and (\ref{eqn: SineAnalyticalu}) respectively,
\begin{align}
  \bar{N}_{\mathrm{sound}}(x,0)&=\bar{N}^{eq}+A_0 \sin\left(\frac{2\tilde{\pi} x}{L}\right),\label{eqn: sine density}\\
  u_{\mathrm{sound}}(x,0)&=0.
\end{align}

\begin{figure}
    \centering
    \includegraphics[width=\columnwidth,clip=true]{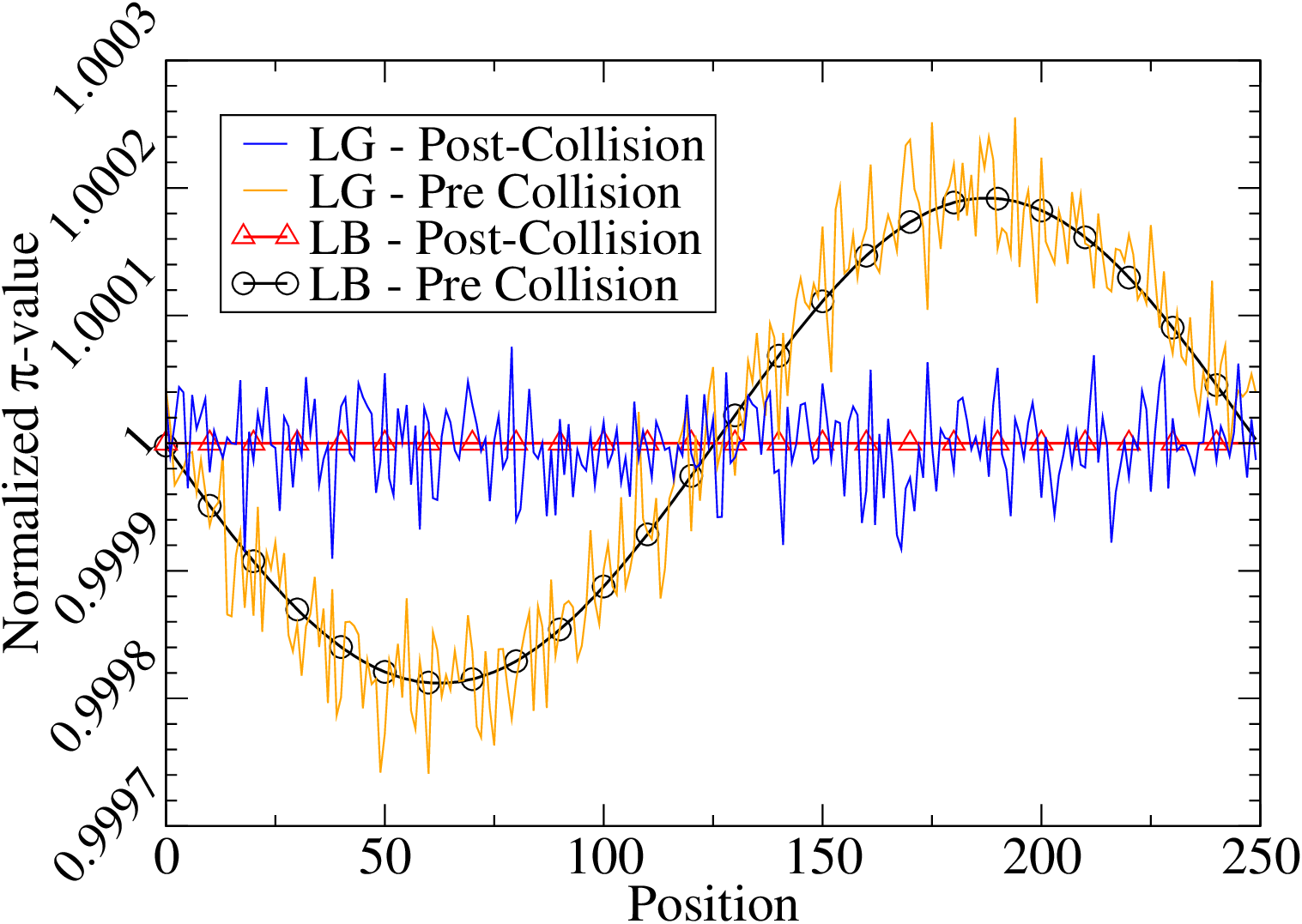}
    \caption{Normalized pre- and post-collision $\pi$ values of a decaying sound wave. The lattice Boltzmann system was initialized using Eq. (\ref{eqn: sine density}) with $\bar{N}^{eq}=1,000$ and $L=250$, and the lattice gas used these same values as a mean for its Poisson distributed initialization. After 50 iterations run at $\omega=\omega_{LB}=1$}, the pre- and post-collision $\pi$ values were measured over the course of a single collision. The lattice gas was averaged over 100,000 random seeds. Both were normalized according to the equilibrium average post-collision $\pi$ value from Eq. (\ref{eqn: Pi Average Eq}).
    \label{fig: SinTheta}
\end{figure}

We can observe the effect of a single collision on the system by looking at the difference between the pre- and post-collision averaged $\pi$ values of the system. We then normalize the system with the expected average post-collision $\pi$ value, which is found by using the $f_i^{0}$ from Eq. (\ref{eqn: Entropic Local Value}),
\begin{equation}
    \bar{\pi}^{0}=f_1^{0}+f_{-1}^{0}.
    \label{eqn: Pi Average Eq}
\end{equation}
This formula assumes the system is in an equilibrium state, however it serves as a good approximation even for non-equilibrium systems. The results for the effect of a single collision on the decaying sound wave can be seen in Fig. \ref{fig: SinTheta}, which shows agreement between the average effect of the collision operator for the BGK entropic lattice Boltzmann algorithm from Eq. (\ref{eqn: BGK}) with an equilibrium distribution given by Eq. (\ref{eqn: Entropic Local Value}) and lattice gas algorithm.

To observe the viscosity of the system, we first consider the amplitude decay of the sine wave as it approaches equilibrium. We utilize a method of amplitude extraction discussed by Blommel and Wagner in an appendix to their paper~\cite{blommel2018integer}. Specifically, the amplitude of our sound wave is given by,
\begin{equation}
    A(t)=\frac{\sum_x\sin\left(\frac{2\tilde{\pi} x}{L}\right) \bar{N}(x,t)}{\sum_x\sin^2\left(\frac{2\tilde{\pi} x}{L}\right)}.
\end{equation}
\begin{figure}
    \centering
    \includegraphics[width=\columnwidth,clip=true]{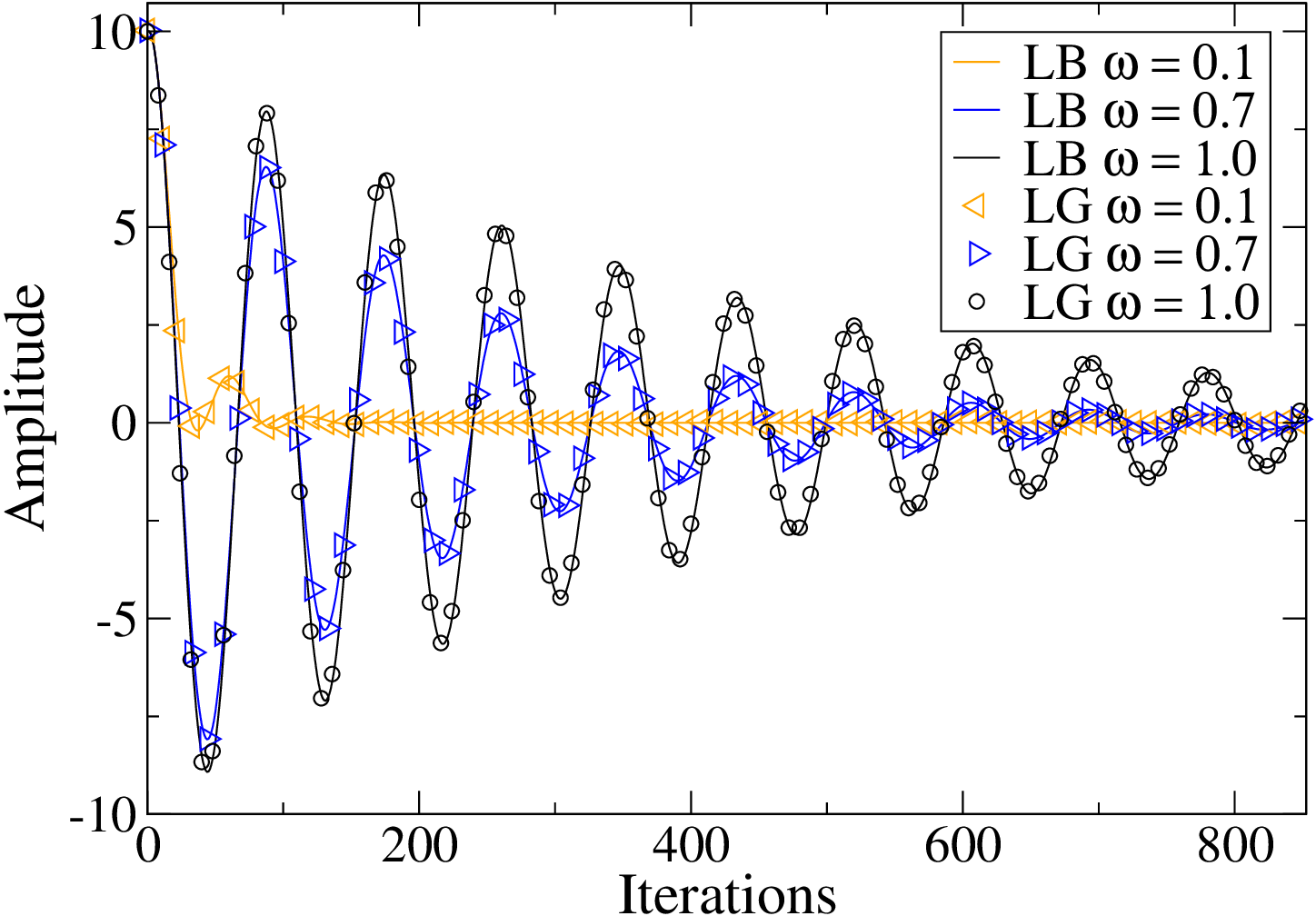}
    \caption{The extracted amplitude evolution of the decaying sound wave for various relaxation times for the lattice gas and lattice Boltzmann simulations over 850 iterations for a lattice size of 50. Systems were initialized otherwise identically to those in Fig. \ref{fig: SinTheta}, and run for $\omega=\omega_{LB}=1.0$, $\omega=\omega_{LB}=0.7$, and $\omega=\omega_{LB}=0.1$.} The amplitude was extracted utilizing Blommel and Wagner's method after each iteration~\cite{blommel2018integer}. Lattice gas systems were averaged over 25,000 random seeds.
    \label{fig: SinDecay}
\end{figure}

Fig. \ref{fig: SinDecay} shows the results of this amplitude extraction for $\omega=\omega_{LB}=1$, $\omega=\omega_{LB}=0.7$, and $\omega=\omega_{LB}=0.1$ over 850 iterations for an average of $\bar{N}^{eq}=1,000$ particles. There is excellent agreement between the lattice gas and the lattice Boltzmann algorithms in this case, despite the fact that the BGK lattice Boltzmann algorithm is not an exact Boltzmann limit for our lattice gas. We can measure the viscosity of the system numerically by extracting the decay rate from the wave amplitude. We can do this by fitting our results from Fig. \ref{fig: SinDecay} to the analytical solution from Eq. (\ref{eqn: SineAnalyticalN}). The amplitude of Eq. (\ref{eqn: SineAnalyticalN}) is given by,
\begin{equation}
    A^{th}(t) = A_0\cos(ft)e^{-\lambda t}.
    \label{eqn: SineDecaySolution}
\end{equation}
The recovered $\lambda$ can then be converted to the kinematic viscosity $\nu$ by inverting Eq. (\ref{eqn: nuAndlambda}).
\begin{figure}
    \centering
    \includegraphics[width=\columnwidth,clip=true]{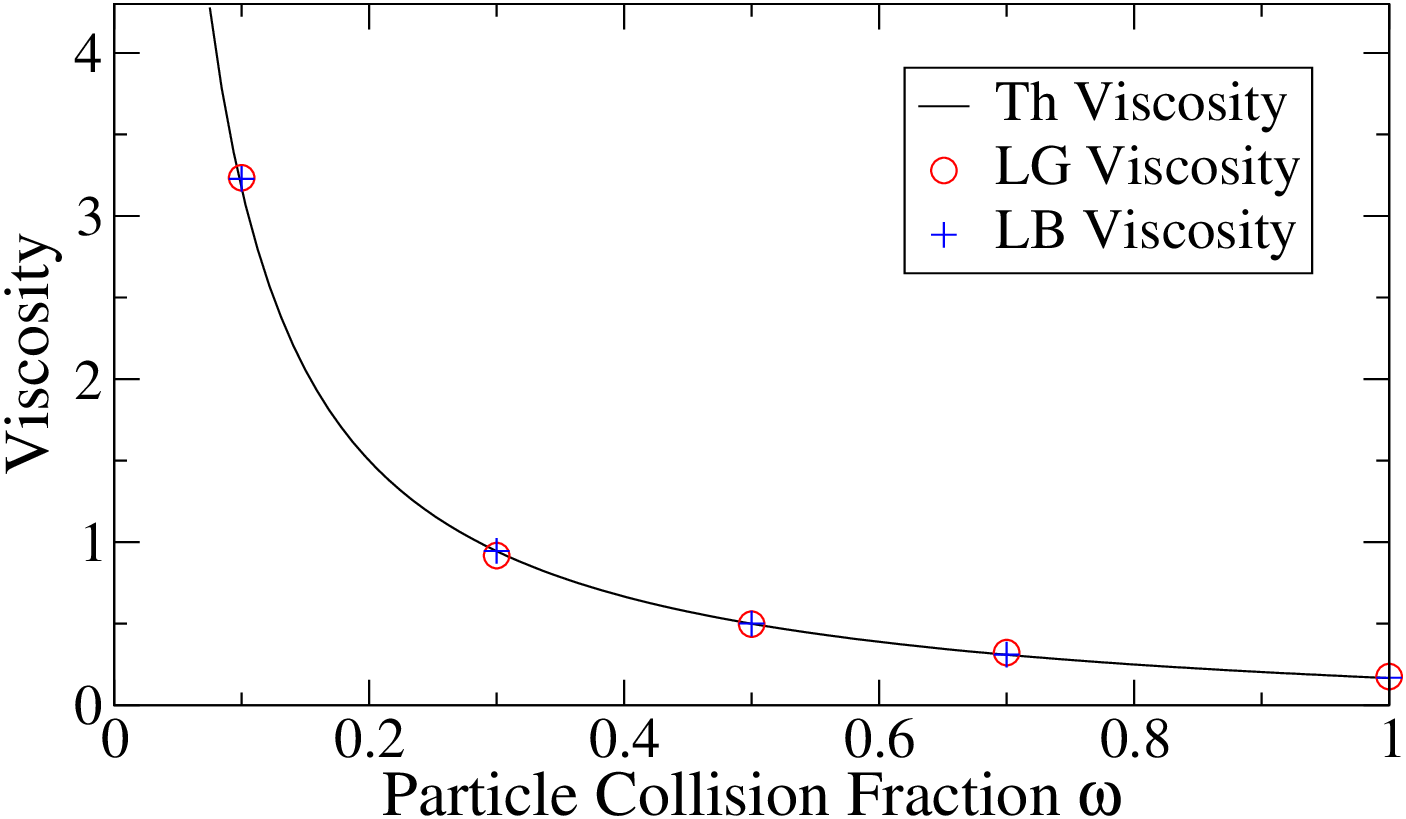}
    \caption{The recovered viscosities of the lattice Boltzmann and lattice Gas corresponding to various relaxation times along with the theoretical scaling from Eq. (\ref{eqn: Kinematic Viscosity}). Each system was initialized and run identically to those in Fig. \ref{fig: SinDecay}. Along with the systems from Fig. \ref{fig: SinDecay}, systems were also run at $\omega=\omega_{LB}=0.3$ and $\omega=\omega_{LB}=0.5$. The $\lambda$ values were found via non-linear curve fitting on xmgrace with the analytical solution from Eq. (\ref{eqn: SineDecaySolution}) and calculated using Eq. (\ref{eqn: nuAndlambda}).}
    \label{fig: ViscosityAndOmega}
\end{figure}
 We can then compare the extracted values of $\nu$ from the lattice Boltzmann and lattice gas systems to the theoretical $\nu_\omega$ value from Eq. (\ref{eqn: Kinematic Viscosity}) for a given value of $\omega$. Fig. \ref{fig: ViscosityAndOmega} shows the results of this comparison for a high density ($\bar{N}^{eq}=1,000$) system.
\begin{figure}
    \centering
    \includegraphics[width=\columnwidth,clip=true]{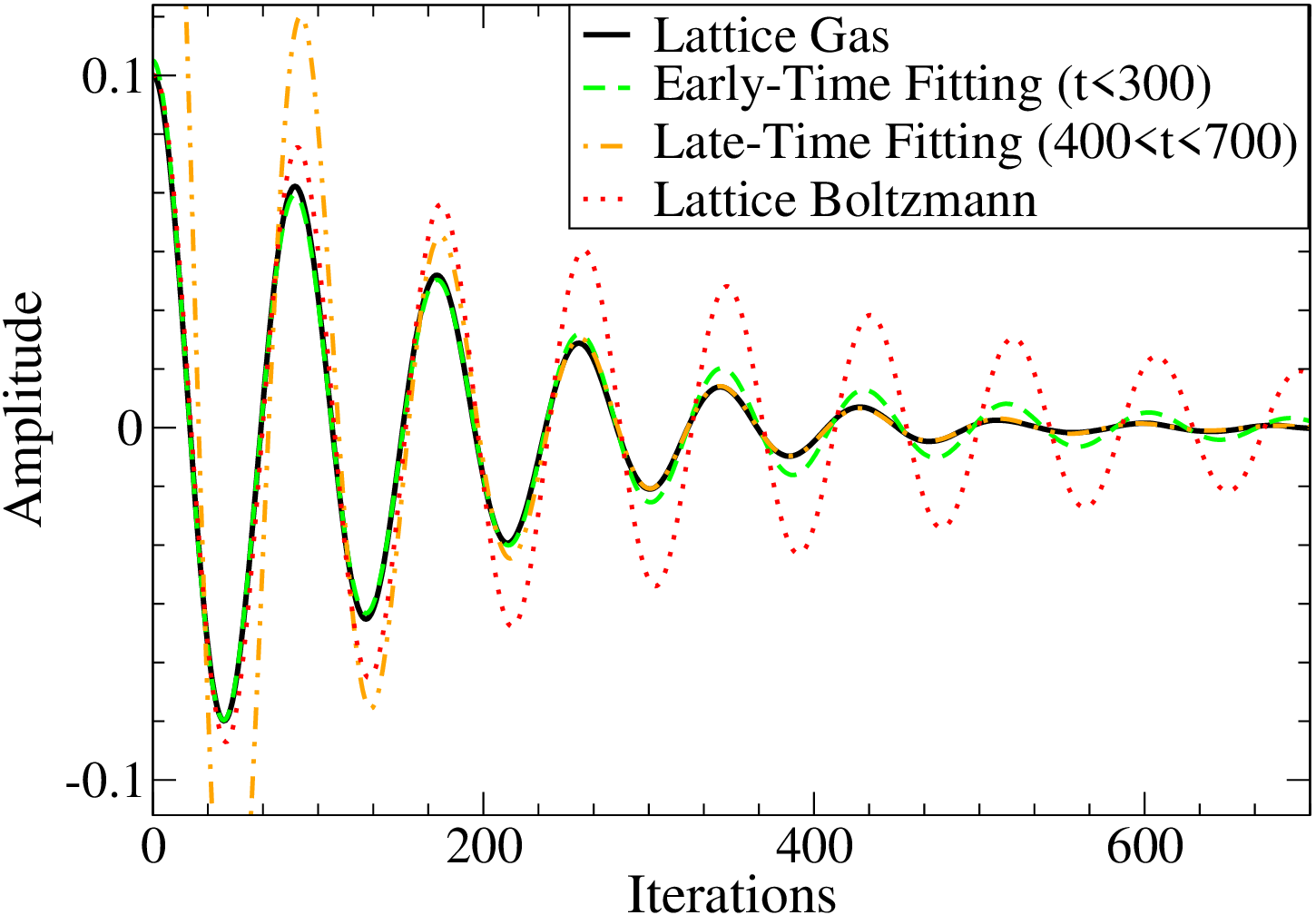}
    \caption{The extracted amplitude of the decaying sound wave for both lattice gas and lattice Boltzmann simulations over 850 iterations run at $\omega=\omega_{LB}=1$. along with two potential fittings at a constant decay rate of for the lattice gas. Simulations are initialized identically to those in Fig. \ref{fig: SinDecay}, except with a $\bar{N}^{eq}$ value of 10. Both fits have been made utilizing non-linear curve fitting on xmgrace assuming Eq. (\ref{eqn: SineDecaySolution}) as the fitting equation, with the fits restricted to Iterations$<300$ (early-time) and $400<$Iterations$<700$ (late-time).}
    \label{fig: MultiDecay}
\end{figure}

There is excellent agreement between the results of the lattice Boltzmann and lattice gas algorithms in this case, as well as between the results from both algorithms and Eq. (\ref{eqn: Kinematic Viscosity}). The results from Figs. \ref{fig: SinDecay} and \ref{fig: ViscosityAndOmega} imply that $\omega$ and $\omega_{LB}$ are equivalent, however this is only true when considering high density systems. For low density systems, the decay rate is non-constant, an effect that can be seen in Fig. \ref{fig: MultiDecay}.

Both the lattice gas and its fittings show a clear distinction from the lattice Boltzmann in the case of this smaller $\bar{N}^{eq}$ value. The fits themselves are also imperfect, due to the fact that Eq. (\ref{eqn: SineDecaySolution}) assumes $\lambda$ to be a constant value. However because the early-time fitting and late-time fitting do not agree with each other, the value of $\lambda$ is not constant with respect to time for the lattice gas. 
\begin{figure}
    \centering
    \includegraphics[width=\columnwidth,clip=true]{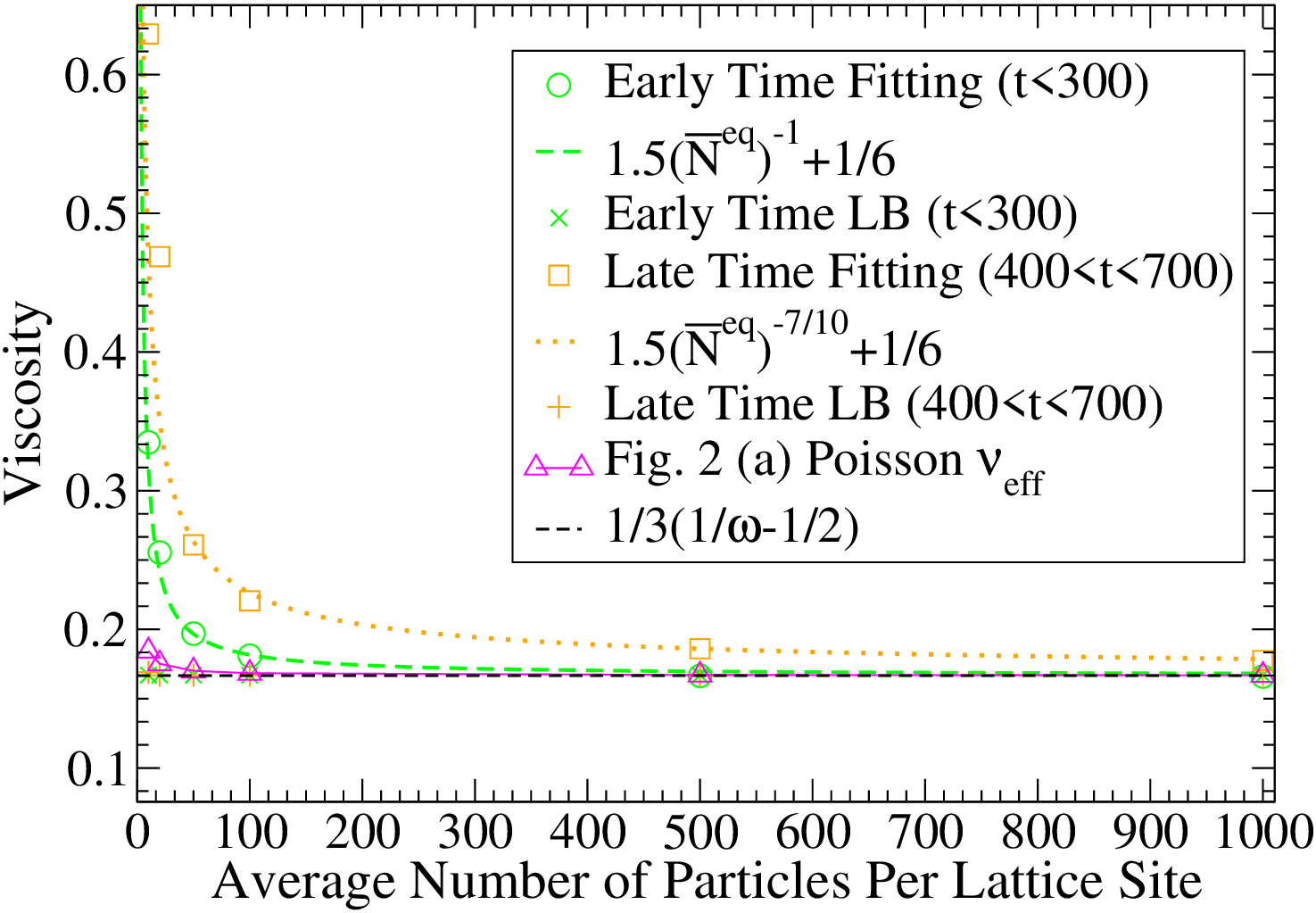}
    \caption{The density dependence of the viscosity of a decaying sound wave for lattice Boltzmann and lattice gas systems. See text for details.}
    \label{fig: densityBased}
\end{figure}

By extracting the decay rate and calculating the viscosity using Eq. (\ref{eqn: nuAndlambda}), as was done in Fig. \ref{fig: ViscosityAndOmega}, and repeating the process for various values of $\bar{N}^{eq}$, we can visualize the scaling of the lattice gas's viscosity with respect to density. This is shown in Fig \ref{fig: densityBased}.

We start by initializing the systems identically to those of Fig. \ref{fig: SinDecay}, but with varying initial values of $\bar{N}^{eq}$. All systems were then run with $\omega=1$ for 1,000 timesteps, averaged over 100,000 random seeds per density. The viscosities were calculated identically to those in Fig \ref{fig: ViscosityAndOmega}. Two sets of fitting parameters were used, corresponding to the early-time and late-time fittings in Fig. \ref{fig: MultiDecay}. We then repeated this process for the BGK entropic lattice Boltzmann algorithm for $\omega_{LB}=1$.

We also considered the collision operator for an uncorrelated Poisson distribution. Since the slope of the collision operator from Fig. \ref{fig: ColOpLinesUnscaled} (a) is slightly different to that of the BGK collision operator, we can obtain a different effective $\omega$ value with which to calculate the viscosity. However, as we will see below, this small correction does not explain the much larger deviations measured in our simulation.

There is a clear deviation from the theoretical viscosity for lower densities in both the early-time and late-time fittings for the lattice gas. The lattice gas's late-time fittings in particular show increased deviation from the theoretical viscosity value even for higher densities. The lattice Boltzmann agrees well with the theoretical viscosity value for both early-time and late-time fittings. This reinforces our findings from Fig. \ref{fig: MultiDecay} that there is a difference in the non-equilibrium behavior of the lattice Boltzmann and lattice gas algorithms that leads to this non-constant decay rate. However as the density increases beyond approximately $\bar{N}^{eq}=1,000$, the difference becomes small, particularly for early-time fittings.

Both lattice gas fittings having a more extreme viscosity scaling compared to that of the collision operator alone implies that the system is not perfectly uncorrelated as it relaxes to equilibrium. An example of this effect was already shown in Fig. \ref{fig: ColOpLinesUnscaled} (a). The inconsistency in the decay rates at low densities indicates that there is some underlying physical effect that leads to a buildup of correlations that decrease the efficiency of the collision operator leading to a smaller effective collision probability. Studying this effect in detail, however, is outside of the scope of this paper, and will be left to future work.

\subsection{The SOD Shock Tube}
The SOD shock tube is defined by an initial state with two regions in equilibrium with differing initial average densities, but identical initial velocities~\cite{SOD1978Survey}. Due to the step discontinuity between the two regions, the SOD shock tube is a Riemann problem~\cite{Toro2009Riemann}. This particular Riemann problem has been used as a basic test case for a variety of hydrodynamic simulation methods~\cite{SOD1978Survey, Ge-Cheng1999Comparative, Loubere2014MOOD, Feng2010Compressible}. 

In our case, we define the initial distribution of average densities as,
\begin{equation}
    \bar{N}_{\mathrm{shock}}(x,0)=
    \begin{cases} 
      N_{l} & x<\frac{L}{2}\\
      N_{r} & \frac{L}{2}\leq x\\
   \end{cases}
   \label{eqn: SOD Density Formula}
\end{equation}
with $N_{l}$ and $N_{r}$ being the average densities of the two initial regions. The initial $\tilde{f}_i$ values for the shock tube can then be written using Eq. (\ref{eqn: Entropic Local Value}), as both regions should initially be in equilibrium internally. The boundary conditions typically include reflective boundary conditions on both ends of the tube, however we found that a mirrored shock wave system allowed us to use periodic boundary conditions, though only one side of the shock wave will be shown due to their symmetry.

The analytical solution for the SOD shock tube for zero viscosity is relatively well studied as a solution to the Euler equations~\cite{Toro2009Riemann}, however our collision operator restricts the system to an isothermal average state. Negro \textit{et al.} derived an analytical solution for the isothermal inviscid SOD shock tube system~\cite{Negro2019Comparison}. For a rightwards moving shock ($N_l>N_r$) Negro's analytical solutions for density and velocity can be written as,
\begin{equation}
    \bar{N}_{\mathrm{shock}}(x,t)=
    \begin{cases}
        N_{l} & x<x_r\\
        N_{l}\exp\left(\frac{x_0-x}{tc_s}+1\right) & x_r\leq x<x_c\\
        N_{l}\exp(-\zeta) & x_c\leq x<x_s\\
        N_{r} & x\geq x_s,\\
    \end{cases}
    \label{eqn: SOD N Analytical}
\end{equation}
and,
\begin{equation}
    u_{\mathrm{shock}}(x,t)=
    \begin{cases}
        0 & x<x_r\\
         \frac{x-x_{0}}{t}+c_s & x_r\leq x<x_c\\
         \zeta c_s & x_c\leq x<x_s\\
        0 & x\geq x_s,
    \end{cases}
    \label{eqn: SOD u Analytical}
\end{equation}
with the assumption that $u(x,0)=0$ for all $x$. $x_s$ is the position of the shock wave front, and is given by,
\begin{equation}
    x_s=\frac{\zeta c_st}{1-\frac{N_{r}}{N_{l}}\exp(\zeta)}+x_0.\\
\end{equation}
We can also find $x_c$, the start of the central density plateau, and $x_r$, the start of the rarefaction fan,
\begin{align}
    x_c&=c_s(\zeta-1)t+x_{0},\\
    x_r&=x_0-c_st.
\end{align}
$x_0$ is the initial position of the discontinuity between regions, and for our initial conditions is given by $x_0=L/2$. $\zeta$ is a dimensionless quantity that can be found by solving,
\begin{equation}
    2+\zeta^2-\frac{N_{l}}{N_{r}}\exp(-\zeta)-\frac{N_{r}}{N_{l}}\exp(\zeta)=0.
\end{equation}
In a finite system, this analytical solution only remains valid until the end of the boundary is reached by the wave in a system with reflective boundary conditions, or until the two shock wave fronts meet for our mirrored system with periodic boundary conditions.
\begin{figure}
    \centering
    \includegraphics[width=\columnwidth,clip=true]{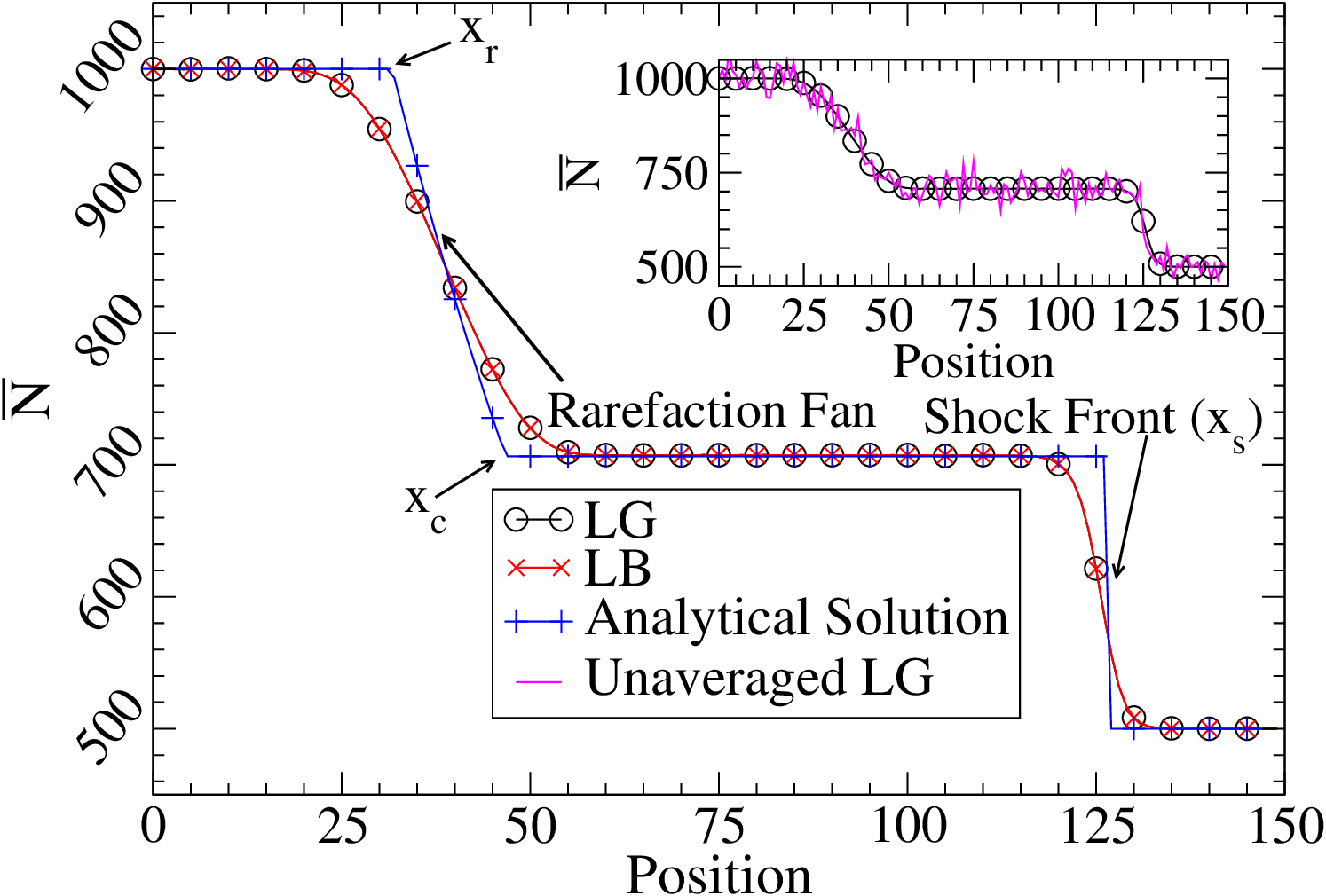}\\(a)\\
    \includegraphics[width=\columnwidth,clip=true]{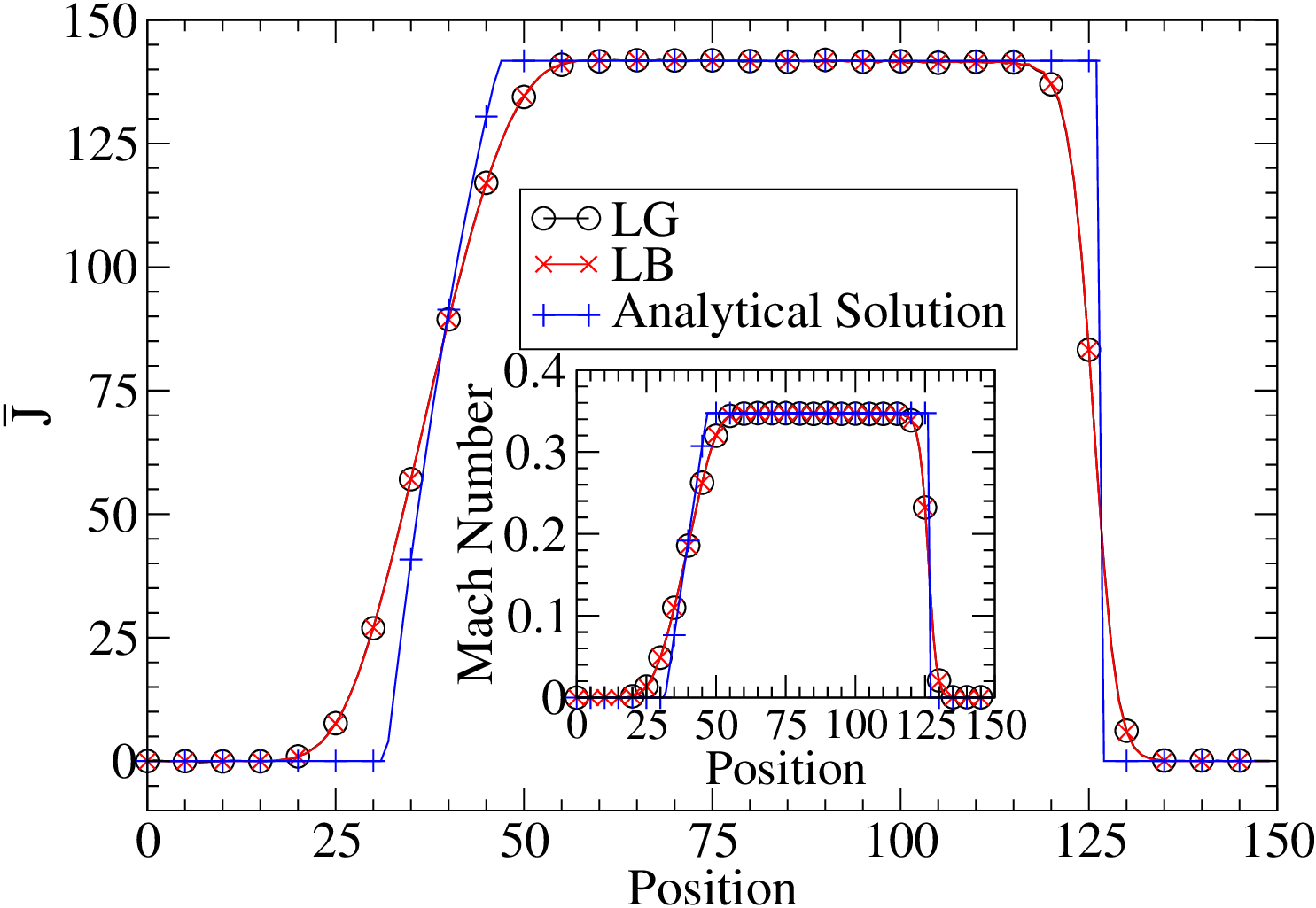}\\(b)
    \caption{The $\bar{N}$ (a) and $\bar{J}$ (b) moment graphs of the isothermal SOD shock tube at 75 iterations for $\omega=\omega_{LB}=1$ comparing the lattice gas results averaged over 25,000 random seeds and the lattice Boltzmann results. The system was initialized with a size of 150 lattice sites, and $\bar{N}=\rho$ values of $N_{l}=1,000$ and $N_{r}=500$, along with a mirrored shock wave on the opposite side of the periodic boundary. The lattice gas $n_i$ values are assigned utilizing a Poisson distribution around Eq. (\ref{eqn: Entropic Local Value}), with the $\bar{N}_{\mathrm{shock}}$ values specified previously and $u_{\mathrm{shock}}(x,0)=0$. (a) includes the analytical solution from Eq. (\ref{eqn: SOD N Analytical}). The analytical solution in (b) is the product of Eqs. (\ref{eqn: SOD N Analytical}) and (\ref{eqn: SOD u Analytical}). (a) also includes labels for various features of the shock wave, and an inset that compares a single seed run of the lattice gas to the averaged lattice gas in the main body of (a). (b) also includes an inset that displays the Mach number, calculated using Eq. (\ref{eqn: Mach Number}).}
    \label{fig: SODTest}
\end{figure}

Fig. \ref{fig: SODTest} shows the equivalence between the lattice Boltzmann and lattice gas simulations with the analytical solution after evolution. Specifically, it considers the ensemble averaged values of the $N$ and $J$ moments: $\bar{N}$ and $\bar{J}$ from Eqs. (\ref{eqn: BarN}) and (\ref{eqn: BarJ}). $\bar{N}$ can be compared directly with $\bar{N}_\mathrm{shock}$ from the analytical solution in Eq. (\ref{eqn: SOD N Analytical}), and $\bar{J}$ can be compared to the analytical solutions in Eqs. (\ref{eqn: SOD N Analytical}) and (\ref{eqn: SOD u Analytical}) multiplied together, since $\bar{J}=u\bar{N}$. Fig. \ref{fig: SODTest} (a) also labels the features of the SOD shock wave, including the rarefaction fan, the shock wave front, and the positions $x_r$, $x_c$, and $x_s$, and includes an inset that shows the fluctuating nature of each run of the lattice gas. This inset shows the result of a run with no averaging, compared to our ensemble average lattice gas used in the main body of Fig \ref{fig: SODTest} (a). Fig. \ref{fig: SODTest} (b) includes an inset for the Mach number at each position along the shock wave. Here we calculate the Mach number assuming a constant speed of sound on our lattice given by Eq. (\ref{eqn: ThetaSoundSpeed}),
\begin{equation}
    M=\frac{u}{c_s},
    \label{eqn: Mach Number}
\end{equation}
however this only serves as an approximation, as $c_s$ is potentially velocity dependent, as was discussed briefly earlier.

The averaged $N$ (Fig. \ref{fig: SODTest} (a)) and $J$ (Fig. \ref{fig: SODTest} (b)) moments for the lattice gas algorithms show an excellent agreement with those of the lattice Boltzmann algorithm. They also show decent agreement with their respective analytical solutions for the inviscid SOD shock tube. There are some small deviations between both sets of simulated data and the analytical solution due to the viscosity not being low enough to recover the piecewise nature of the analytical solutions. For lattice Boltzmann simulations, it is possible to achieve closer agreement by using $\omega_{LB}>1$, however, as mentioned above, collision probabilities greater than one are not yet an option for the hydrodynamic lattice gas algorithm.

While gathering data for Fig. \ref{fig: SODTest}, we also considered the initial fluctuations in the moments, as we wished to see whether initial system fluctuations had any effect on the collision operator the way correlations were shown to in Fig. \ref{fig: ColOpLinesUnscaled} (a). To this end, simulations with two different initial particle distributions were run. The first of these distributions is a Poisson distribution of the $n_i$ values given by Eq. (\ref{eqn: Poisson ni}), with Eq. (\ref{eqn: SOD Density Formula}) and $N_{l}=1,000, N_{r}=500$ for its initial $\bar{N}$ values. The second distribution was a flat distribution given by a delta function around the entropic lattice Boltzmann equilibrium distribution,
\begin{equation}
    P_{SOD\delta}(\{n_i\}_l,x,0)=\prod_i\delta\left(n_i-f_i^{0}\left(\bar{N}(x,0),u(x,0)\right)\right).
    \label{eqn: Delta P}
\end{equation}
$P_{SOD\delta}(\{n_i\}_l,0)$ also has the restriction that $f_i^\mathrm{0}\in \mathbb{N}_0$. As the averaged values of the moments resultant from both of these initial probability distributions are indistinguishable to the order of $10^{-4}$, which is within the bounds of the averaging of our data set, only the results from the Poisson distributed system are shown for the lattice gas in Fig. \ref{fig: SODTest}.

\begin{figure}
    \centering
    \includegraphics[width=\columnwidth,clip=true]{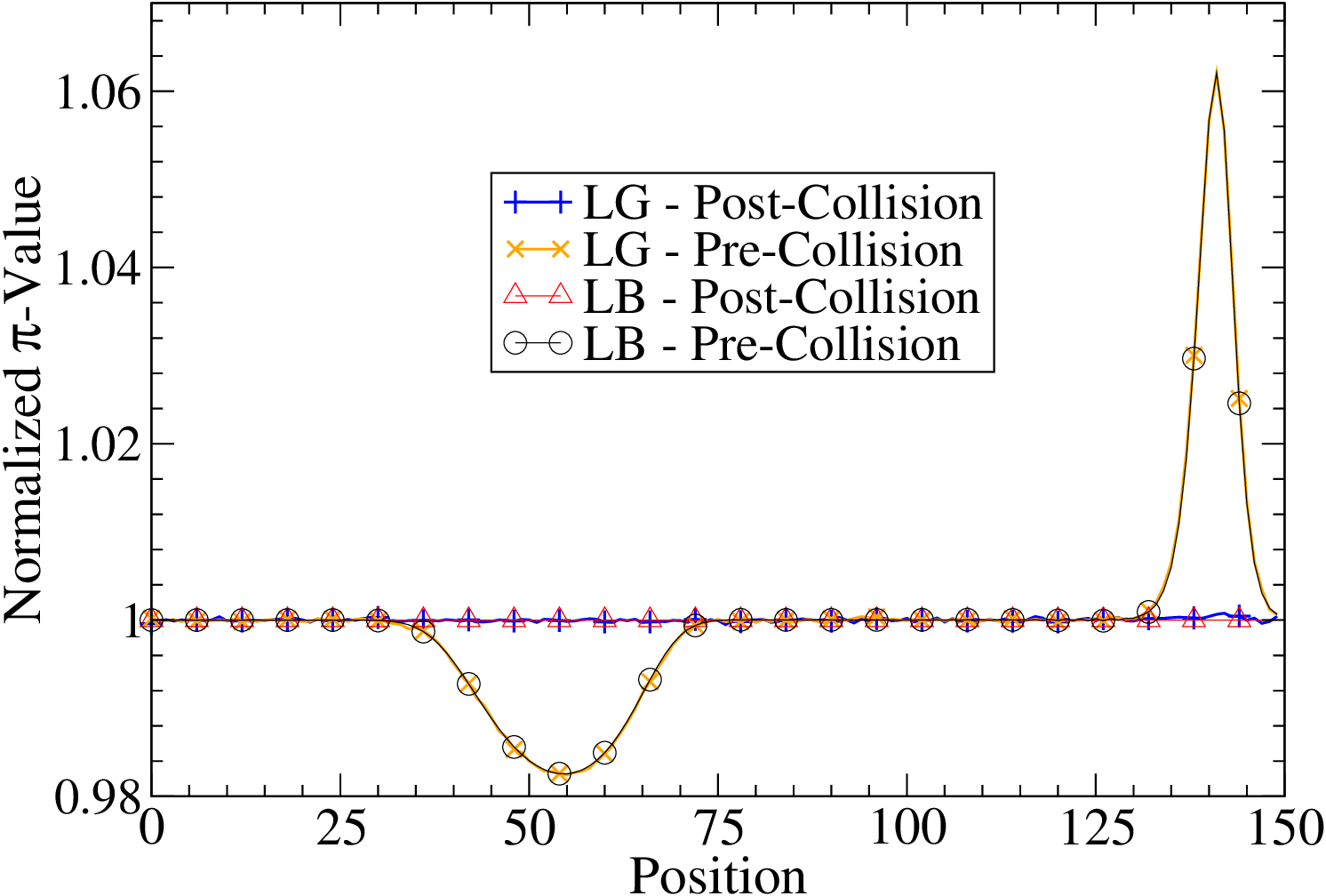}
    \caption{Normalized pre- and post-collision $\pi$ values of the SOD Shock Tube. The systems were initialized identically to those in Fig. \ref{fig: SODTest}. After 75 iterations, another collision was run for each system and the pre- and post-collision $\pi$ values were measured for that extra collision. The lattice gas was averaged over 100,000 random seeds. Both were normalized according to the average post-collision $\pi$ value from Eq. (\ref{eqn: Pi Average Eq}).}
    \label{fig: SODTheta}
\end{figure}

Since the system dynamics are driven by the collision operator we examined the pre- and post-collision $\bar{\pi}$ values for a single collision, similarly to what was done in Fig. \ref{fig: SinTheta}. We can see in Fig. \ref{fig: SODTheta} that these pre- and post-collision results show good agreement between the lattice Boltzmann and lattice gas algorithms. We can also note the fact that for the parts of this system that are in equilibrium, the average collision operator is identical between the pre- and post-collision states. These equilibrium regions include both those that were initialized in equilibrium and have yet to be affected by the shock behavior, as well as the bulk of the shock wave itself ($x_c\leq x<x_s$). The other regions of the wave, including the rarefaction fan and the shock front, show highly non-equilibrium behavior. The shock front shows an increase in the average pre-collision $\pi$ value, and the rarefaction fan shows a decrease in the average pre-collision $\pi$ value.

\subsection{Deviation from the Analytical Solution}
Looking at how the lattice Boltzmann and the lattice gas algorithms behave under extreme initial conditions for the SOD shock tube system allows us to compare how both algorithms deviate from the analytical solution. As the analytical solution found by Negro is a solution to the Euler equations, it is necessarily inviscid, and thus assumes that the viscosity of the system is zero~\cite{Negro2019Comparison}. Therefore, the analytical solution for the SOD shock tube should differ from the data from both algorithms to some degree for any finite value of $\omega<2$, as can be seen in Fig. \ref{fig: SODTest}.
\begin{figure}
    \centering
    \includegraphics[width=\columnwidth]{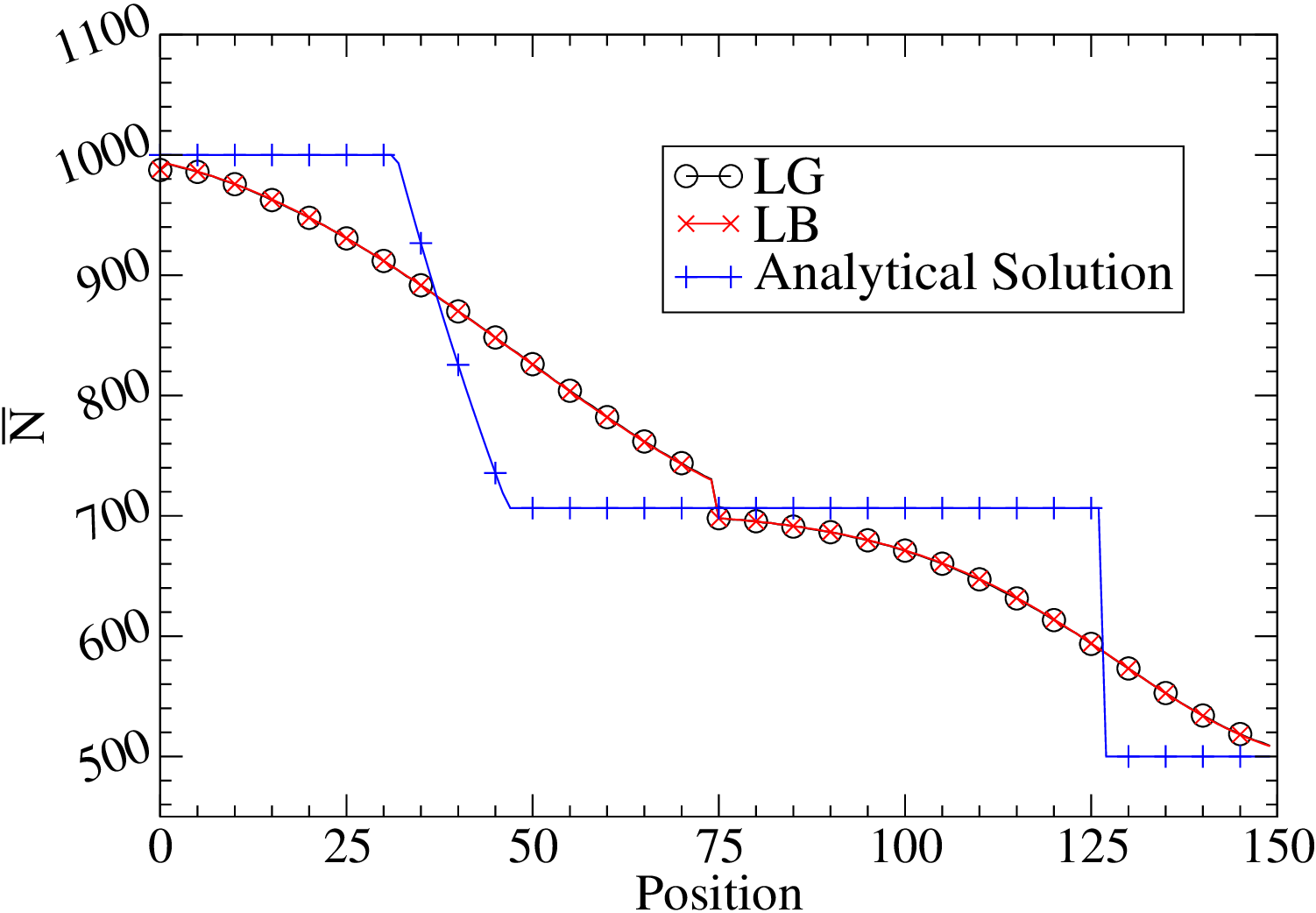}\\(a)\\
    \includegraphics[width=\columnwidth]{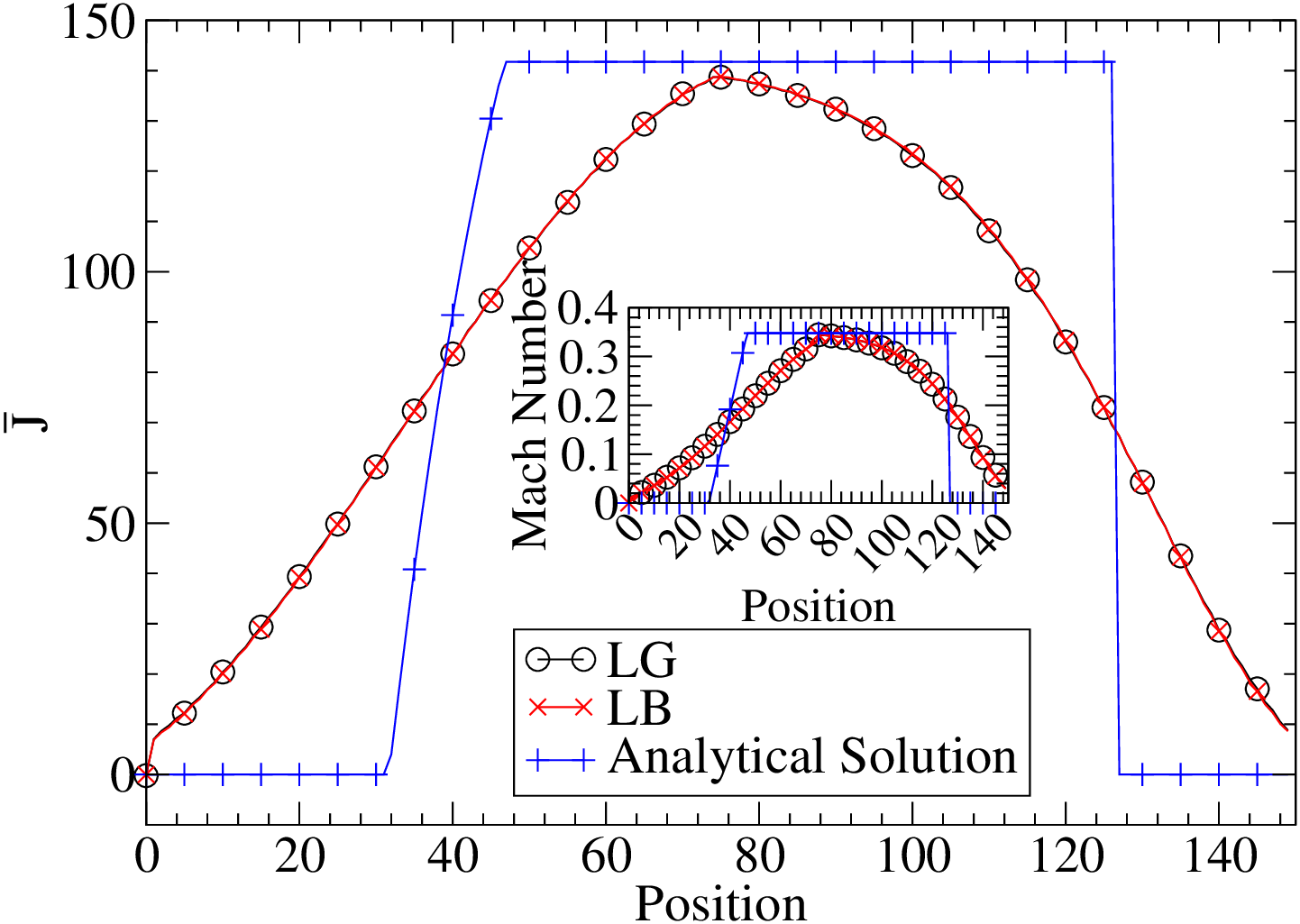}\\(b)\\
    \includegraphics[width=\columnwidth]{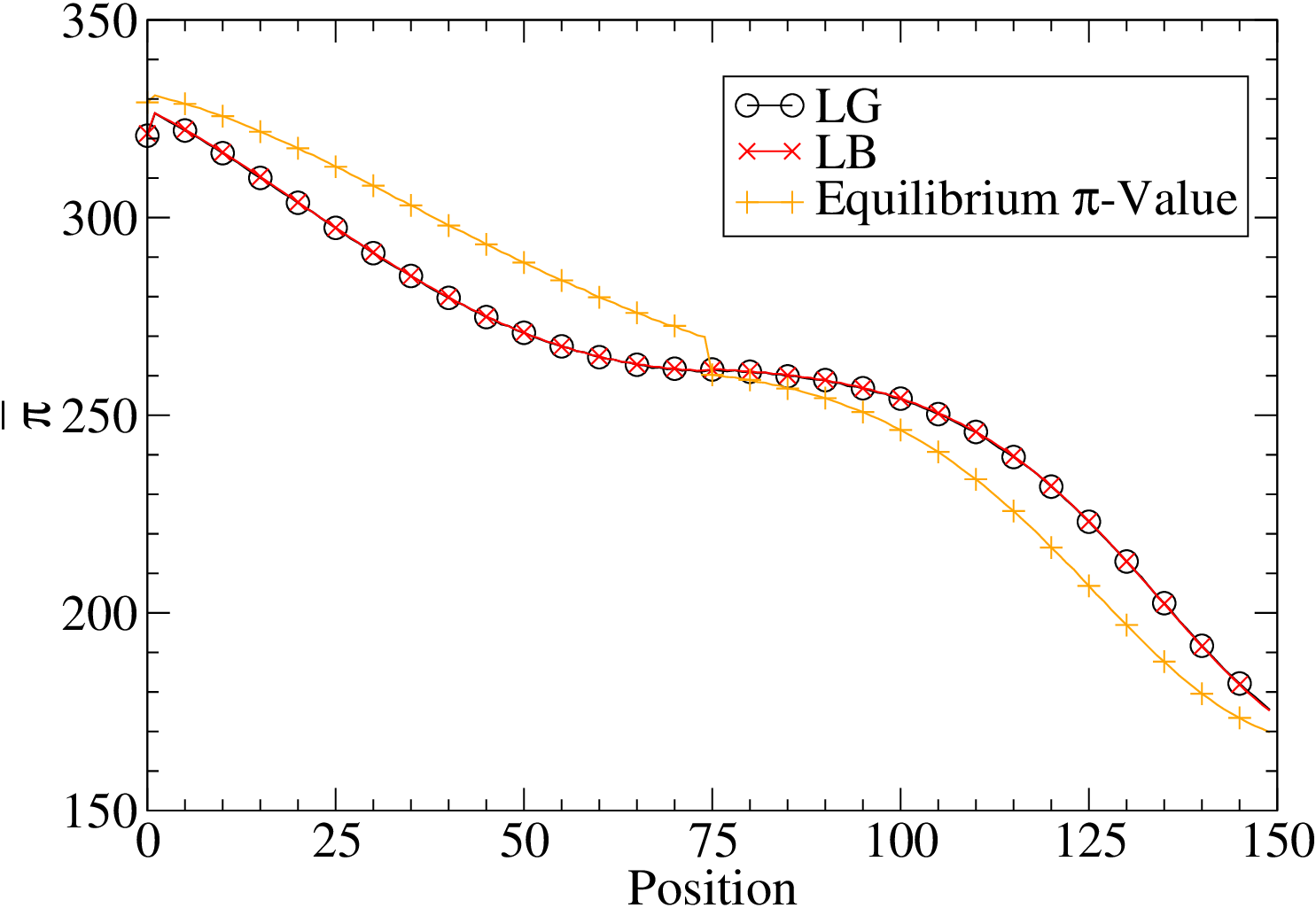}\\(c)
    \caption{The $\bar{N}$ (a), $\bar{J}$ (b), and $\bar{\pi}$ (c) moment graphs of the isothermal SOD shock tube at 75 iterations for $\omega=\omega_{LB}=0.1$ comparing the lattice gas results averaged over 25,000 random seeds and the lattice Boltzmann results. All systems were initialized and run identically to those in Fig. \ref{fig: SODTest} aside from the $\omega$ and $\omega_{LB}$ values. (a) and (b) include identical analytical solutions to those in Fig \ref{fig: SODTest}. (b) also includes an inset for the Mach number calculated using Eq. (\ref{eqn: Mach Number}). (c) includes the equilibrium $\bar{\pi}^0$ given by Eq. (\ref{eqn: Pi Average Eq}).}
    \label{fig: SODLowOmega}
\end{figure}

Fig. \ref{fig: SODLowOmega} compares the lattice Boltzmann and lattice gas results for a system with a much higher viscosity, being run at $\omega=\omega_{LB}=0.1$. Specifically it compares the $\bar{N}$ (Fig. \ref{fig: SODLowOmega} (a)), $\bar{J}$ (Fig. \ref{fig: SODLowOmega} (b)), and post-collision $\bar{\pi}$ (Fig. \ref{fig: SODLowOmega} (c)) moments to the applicable analytical solutions in Fig. \ref{fig: SODLowOmega} (a) and (b), and the equilibrium average $\bar{\pi}$ value in Fig. \ref{fig: SODLowOmega} (c).

Though the agreement with the analytical solutions for $\bar{N}$ and $\bar{J}$ is poor due to the large difference in viscosity between the simulations and the analytical solution, the lattice gas and lattice Boltzmann results show excellent agreement with each other. Similarly, the fact that $\omega=\omega_{LB}=0.1$ means that the system will not fully relax to equilibrium in one timestep, meaning that the post-collision $\bar{\pi}$ values from the lattice Boltzmann and lattice gas algorithms do not recover the equilibrium $\bar{\pi}$ value, however the two simulations agree well with each other. This is important to note, as it implies that even the highly non-equilibrium state that the SOD shock tube represents still has a Boltzmann limit that is very well approximated by the BGK entropic lattice Boltzmann algorithm, something that Fig. \ref{fig: ColOpLinesUnscaled} shows is not guaranteed.

To further test the deviation from the analytical solution of the lattice Boltzmann and lattice gas results, we can consider an extreme initial density disparity between $N_l$ and $N_r$. This leads to a deviation from the analytical solution because as the initial density disparity between the two regions increases, the analytical solution requires $u_\mathrm{shock}$ values that approach or exceed $1$. Because $1$ is the maximum possible value of $u$ a lattice Boltzmann or lattice gas system can have, simulations in this range typically show significant deviations. To show this, we consider the extreme density ratio of $20:1$ for $N_l$ and $N_r$. For this particular density ratio, the $u_\mathrm{shock}$ value of the analytical solution is approximately $0.9$ on the central density plateau, as opposed to the analytical solution for the density ratio from Fig. \ref{fig: SODTest}, which has a $u_\mathrm{shock}$ value on the central density plateau of only around $0.2$.
\begin{figure}
    \centering
    \includegraphics[width=\columnwidth]{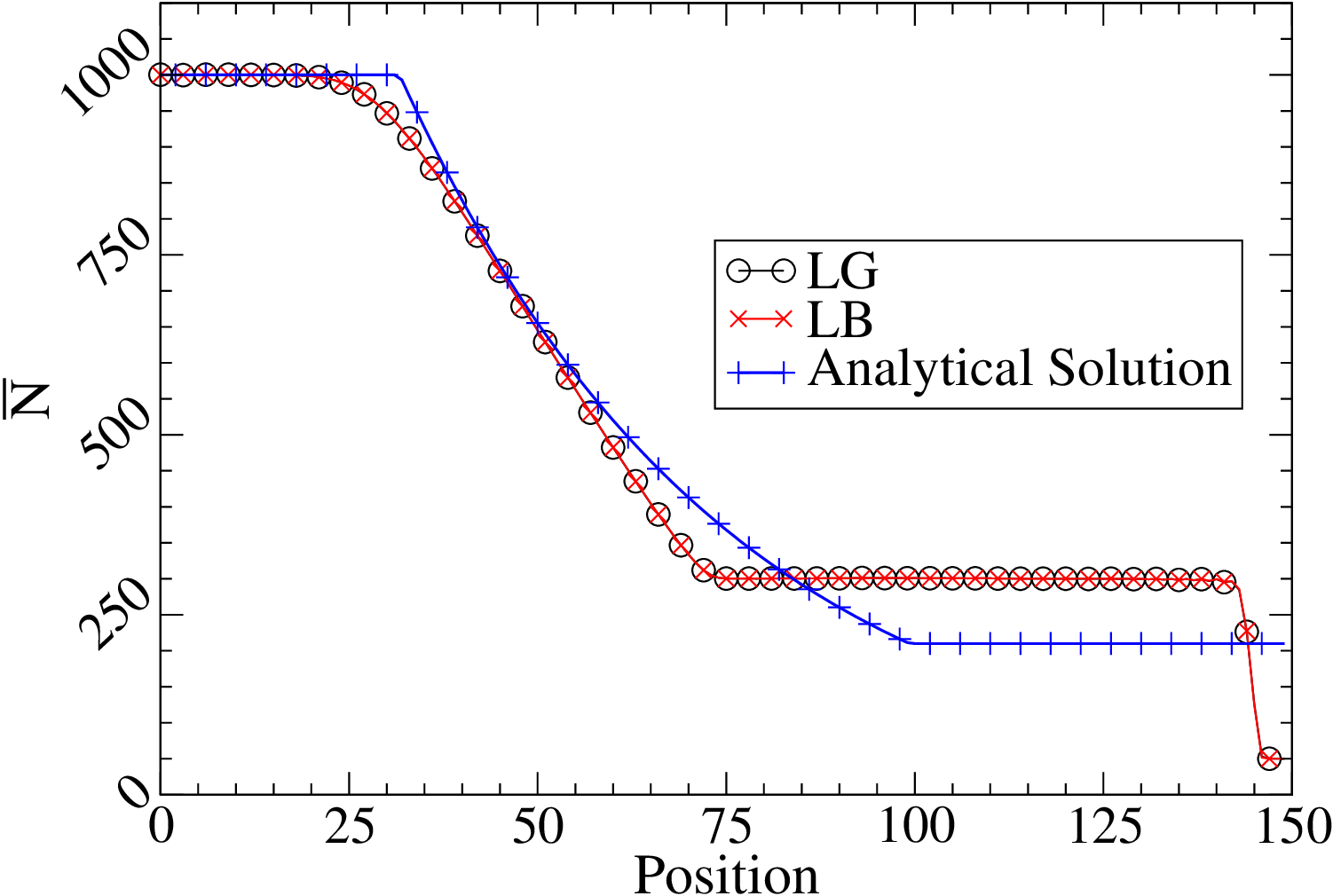}\\(a)\\
    \includegraphics[width=\columnwidth]{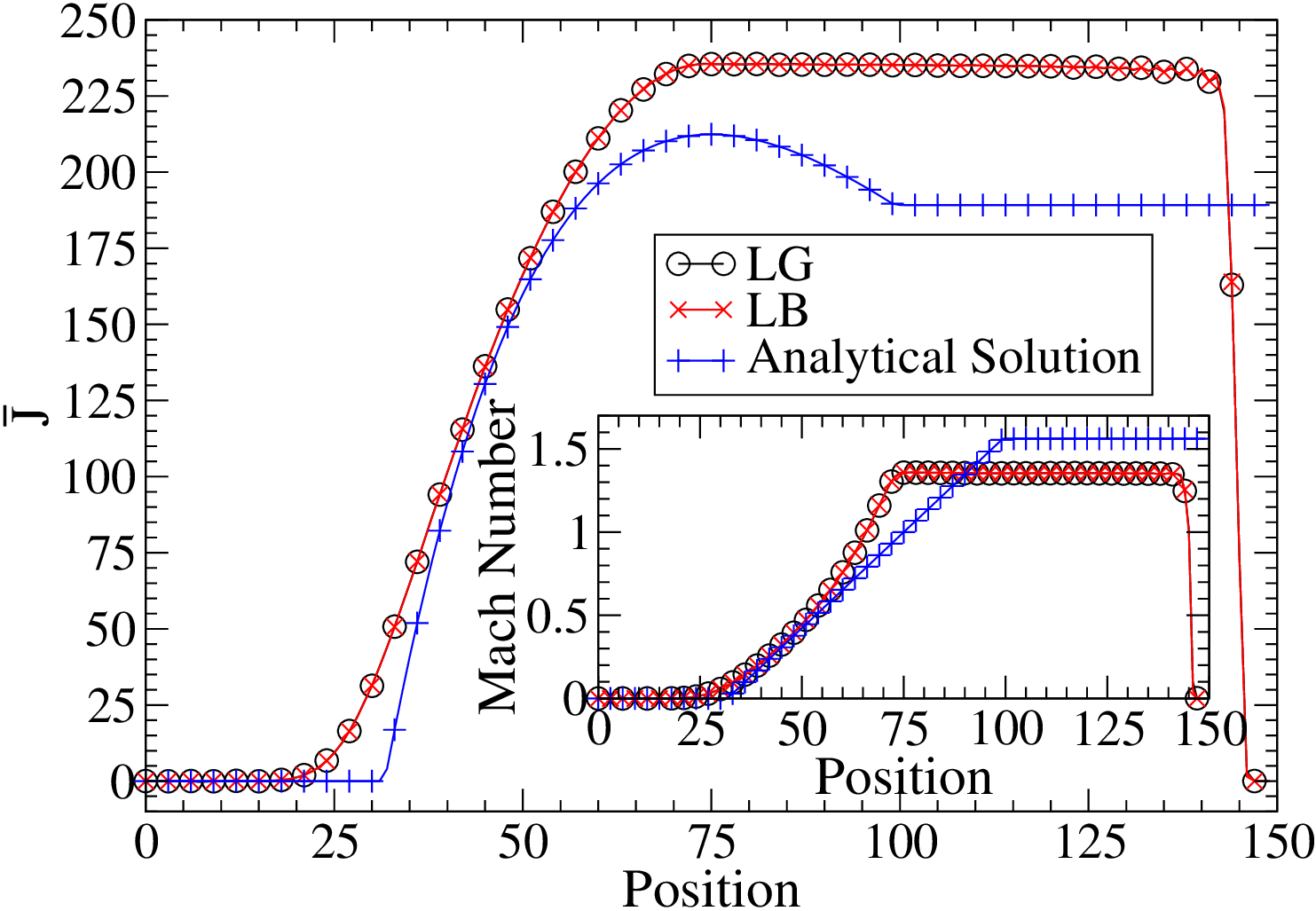}\\(b)
    \caption{The $\bar{N}$ (a) and $\bar{J}$ (b) moment graphs of the isothermal SOD shock tube at 75 iterations for $\omega=\omega_{LB}=1$ comparing the lattice gas results averaged over 25,000 random seeds and the lattice Boltzmann results. The systems were initialized with $\bar{N}$ values of $N_{l}=1,000$ and $N_{r}=50$, but were otherwise initialized and run identically to those in Fig. \ref{fig: SODTest}. Analytical solutions calculated identically to those in Fig. \ref{fig: SODTest}, aside from the different $N_l$ and $N_r$ values used, are included for $\bar{N}$ and $\bar{J}$. (b) also includes an inset for the Mach number calculated using Eq. (\ref{eqn: Mach Number})}
    \label{fig: SODDisp}
\end{figure}

Fig. \ref{fig: SODDisp} shows the results of this increased density disparity for the $\bar{N}$ (Fig. \ref{fig: SODDisp} (a)) and $\bar{J}$ (Fig. \ref{fig: SODDisp} (b)) moments, as well as including the applicable analytical solutions. The agreement between the lattice Boltzmann and lattice gas systems is excellent, as it was in both Fig. \ref{fig: SODTest} and Fig. \ref{fig: SODLowOmega}, however as expected the agreement between the simulation results and the analytical solution is poor. 

Looking at the velocity on the plateau of the wave ($x_c\leq x<x_s$) which is where it is maximal, there is a clear difference between the simulation results and the analytical solution. The $\bar{J}$ value on the plateau corresponds to a $u_\mathrm{shock}$ value of approximately $0.78$ for the lattice Boltzmann and lattice gas simulations. This velocity disparity also explains why the values of $x_c$ and $x_s$ differ between the simulated data and the analytical solutions, as the $\theta$ values for the high velocity state will not be $1/3$, leading to artifacts in the Navier-Stokes equation as was discussed previously~\cite{blommel2018integer}.

\subsection{Non-equilibrium Fluctuations in the Isothermal SOD Shock Tube}
While fluctuations in equilibrium are reasonably well understood, much less is known about non-equilibrium fluctuations. Our integer lattice gas algorithm allows us to examine these non-equilibrium fluctuations. Thus, we consider here the non-equilibrium fluctuations of an isothermal SOD shock wave as a proof of principle.

We limit our investigation of the non-equilibrium fluctuations to the squared standard deviation of a quantity $X$,
\begin{equation}
    \sigma_X^2 = \left\langle (X-\langle X \rangle)^2\right\rangle.
    \label{eqn: Std Dev Fluct}
  \end{equation}
Utilizing the same set of simulations from Fig. \ref{fig: SODTest}, we examined the fluctuation of the moments in the lattice gas, comparing their average non-equilibrium fluctuations to those of global equilibrium. As the equilibrium fluctuations in $N$ and $\pi$ values are given by Eq. (\ref{eqn: Poisson ni}), we normalize the average non-equilibrium fluctuations for $N$ and $\pi$ with $\bar{N}$ and $\bar{\pi}$ respectively. 
  
The equilibrium fluctuations in the $J$ moment do not have the local $\bar{J}$ value as their average, as $J$ is represented by the difference between two Poisson distributed quantities (those being $n_1$ and $n_{-1}$). Therefore, its global equilibrium fluctuations are Skellam distributed~\cite{Skellam1946Difference}. A Skellam distribution has a variance corresponding to the sum of the mean values of the two Poisson distributions that make up the Skellum distribution~\cite{Romani1956Variance}. The average value for our $J$ fluctuations, therefore, becomes $\bar{\pi}$. 
\begin{figure}
    \centering
    \includegraphics[width=\columnwidth]{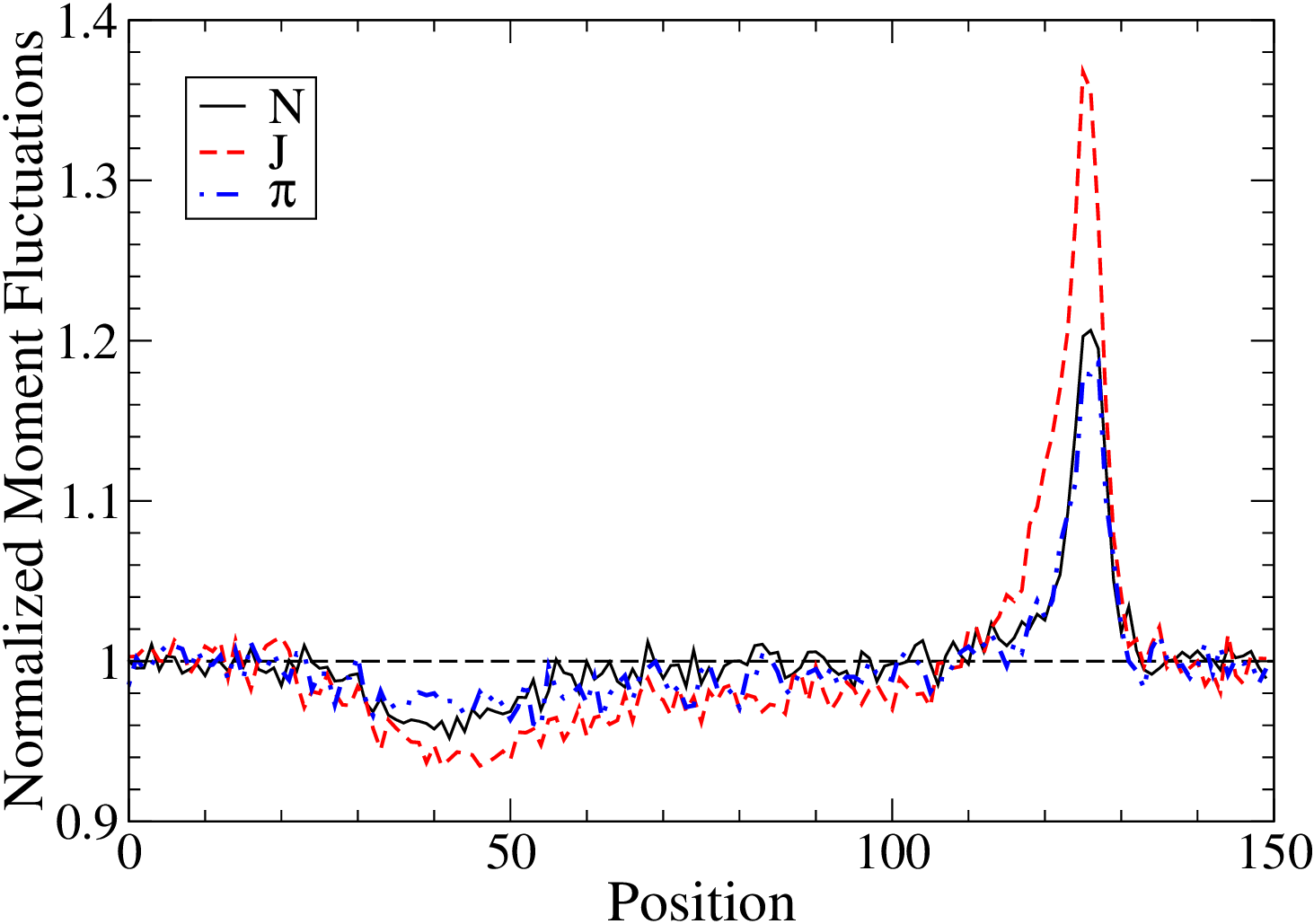}
    \caption{Normalized nonequilibrium fluctuations of the $N$, $J$, and $\pi$ moments after 75 iterations. Initialized identically to the system in Fig. \ref{fig: SODTest}, the system was run for 75 iterations at $\omega=1$ and averaged over 25,000 random seeds. Each moment's variance was then calculated and normalized by dividing by its mean. For the $N$ moment this mean is the local $\bar{N}$ value (Fig. \ref{fig: SODTest}(a)) and for the $J$ and $\pi$ moments this mean is the $\bar{\pi}$ value from Eq. (\ref{eqn: Pi Average Eq}).}
    \label{fig: NJPiFluctNorm}
\end{figure}

The fluctuation results can be seen in Fig. \ref{fig: NJPiFluctNorm}. In the bulk of the shock wave and the regions initialized in equilibrium, the non-equilibrium fluctuations match the equilibrium fluctuations, deviating slightly due to limited averaging. However, in the highly non-equilibrium parts of the shock wave the non-equilibrium fluctuations clearly deviate from the equilibrium fluctuations, with a slight decrease of the fluctuations in the rarefaction fan and a drastic increase around the shock front, mirroring the effects seen in the pre-collision $\pi$ values in Fig. \ref{fig: SODTheta}. While this correlation is very evocative we don't yet understand why it occurs.

The slight decrease in the average $J$ fluctuations in the bulk of the shock wave when compared to the equilibrium fluctuations was noticed as a potential aberration. However when tested in a flat system with an initial advected velocity it was determined to be a product of the $J$ fluctuations taking additional time to reach their equilibrium value rather than a directional bias in the algorithm.

\begin{figure}
    \centering
    \includegraphics[width=\columnwidth,clip=true]{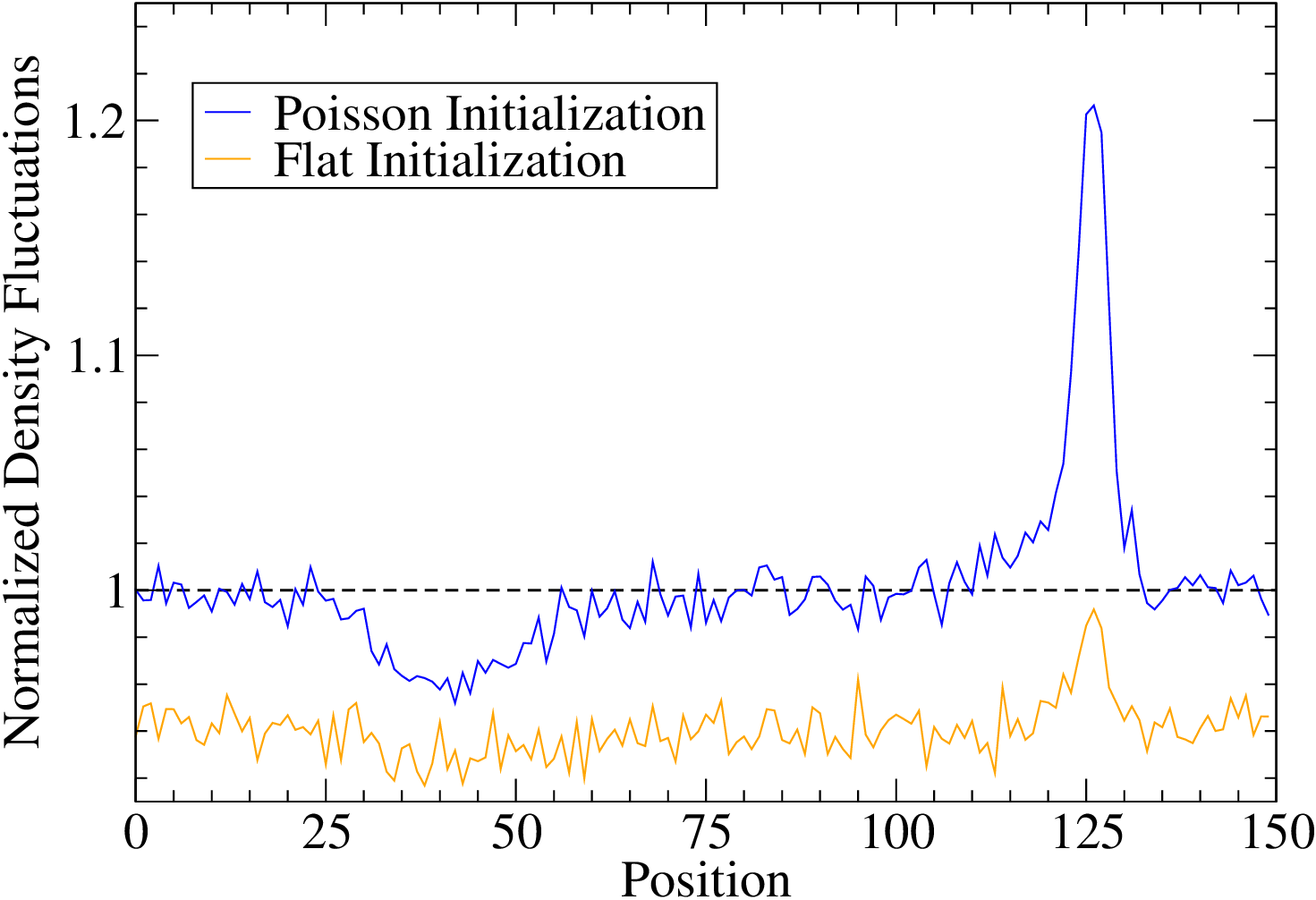}
    \caption{Normalized nonequilibrium $N$ moment fluctuations for Poisson and delta distributed initial fluctuations after 75 iterations. The Poisson distributed system was initialized identically to the system in Fig. \ref{fig: SODTest}, however the flat system was initialized utilizing constant $n_i$ values corresponding to the values of $f^0_i$ as in Eq. (\ref{eqn: Delta P}). The systems were run, and the variances were calculated and normalized identically to those in Fig. \ref{fig: NJPiFluctNorm}.}
    \label{fig: NFluctCompare}
\end{figure}

This delayed timescale for the buildup of fluctuations is also present when the system is initialized with a flat density distribution corresponding to Eq. (\ref{eqn: Delta P}). Since there are no initial fluctuations, all moment fluctuations must build up over time, leading to reduced fluctuations compared to those present in Fig. \ref{fig: NJPiFluctNorm}. These reduced fluctuations can be seen in Fig \ref{fig: NFluctCompare}.

\subsection{Galilean Invariance of the SOD Shock Tube}
As stated previously, the non-Galilean invariance of the collision operator for Boolean lattice gases was a major factor in the widespread adoption of the BGK lattice Boltzmann model~\cite{frisch1987lattice, qian1992lattice}. Due to this, it is important that we show to what extent our lattice gas algorithm maintains Galilean invariance if we seek the sampling lattice gas algorithm to be competitive with the lattice Boltzmann algorithm. 

To numerically test the Galilean invariance of our system, we initialized a set of SOD shock tube systems similarly to those in Fig. \ref{fig: SODTest} for both the lattice Boltzmann and lattice gas algorithms with a lattice size of $L=1,000$ and a net velocity of zero, and a second set of identical systems with a non-zero constant velocity offset $u_g$. After running all simulations (including averaging 2,500 random seeds for each lattice gas system) for 400 iterations at $\omega=\omega_{LB}=1$, we adjusted the coordinate systems for the initial $u_g$ values to take into account the additional distance the simulation would have traveled on average due to the velocity by using,
\begin{equation}
    x'=[x+\lfloor u_gt\rfloor]\bmod[{L}],
    \label{eqn: uAdj}
\end{equation}
which takes into account our system's periodic boundary conditions. $[a]\bmod[b]$ is the modulo operator, returning the positive remainder when $b$ divides $a$. We then calculated the mean squared error in $\bar{N}$ and $u$,
\begin{align}
    \epsilon(\bar{N}(t))=\frac{1}{L}\sum^{L}_{x=0}(\bar{N}(x',t)-\bar{N}_0(x,t))^2,
    \label{eqn: MeanSquaredErrorN}\\
    \epsilon(u(t))=\frac{1}{L}\sum^{L}_{x=0}(u(x',t)-u_g-u_0(x,t))^2,
    \label{eqn: MeanSquaredErroru}
\end{align}
where $\bar{N_0}$ and $u_0$ are the $\bar{N}$ and $u$ data for the $u_g=0$ system.
 
Because we used a mirrored shock wave that propagates opposite the shock waves shown in the previous figures in place of a wall, we can look at the effect of both positive and negative initial values of $u_g$. To do this, the lattice was divided at the center of the shock wave, and the shock moving in the positive x-direction was used to define the error for $u_g>0$, and the shock moving in the negative x-direction was used to define the error for $u_g<0$. Thus, the error for the shock moving in the positive x-direction is higher than the error in the negative x-direction because it is moving at a faster velocity relative to $u_g$.
\begin{figure}
    \centering
    \includegraphics[width=\columnwidth,clip=true]{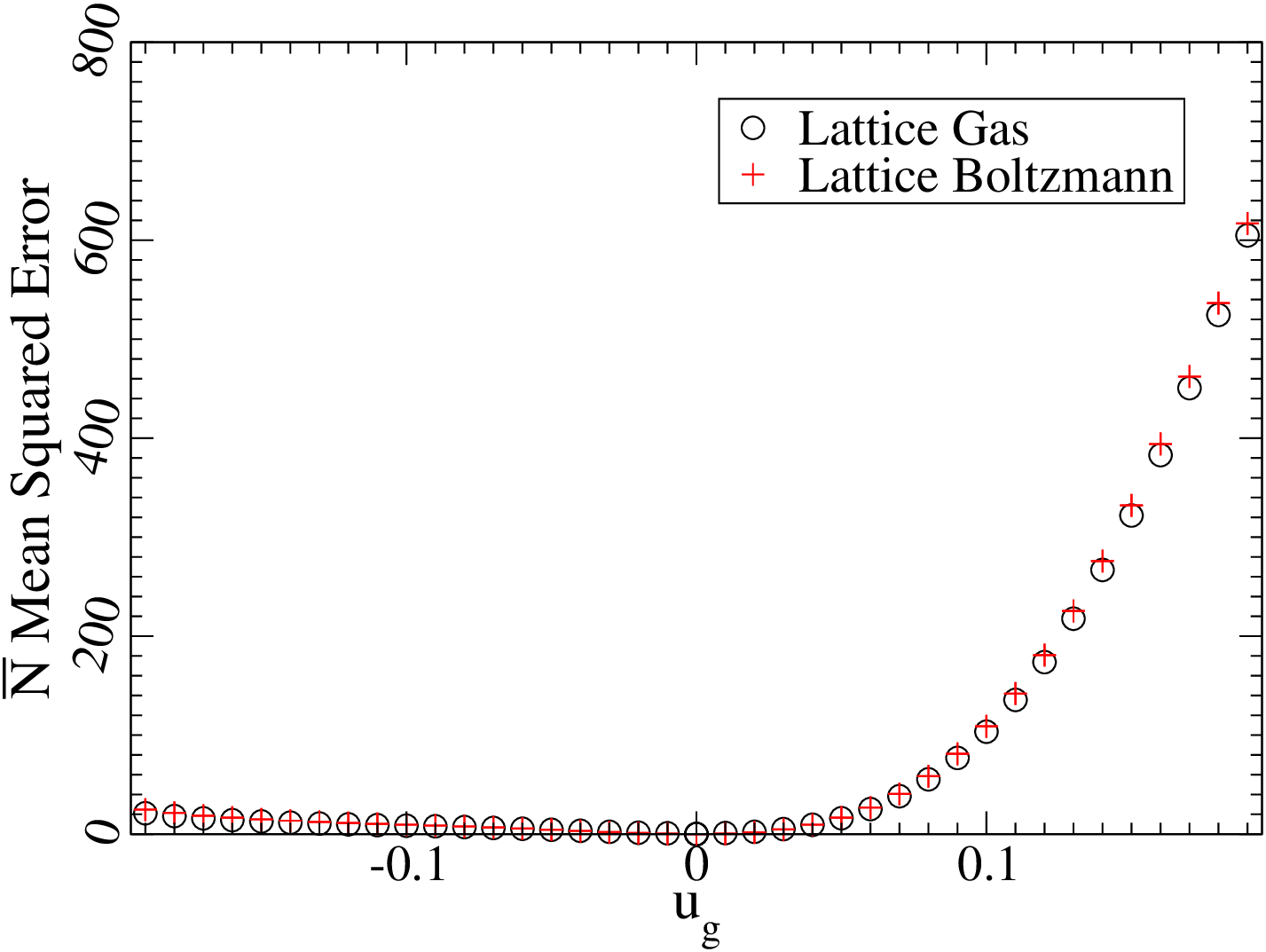}\\(a)\\
    \includegraphics[width=\columnwidth,clip=true]{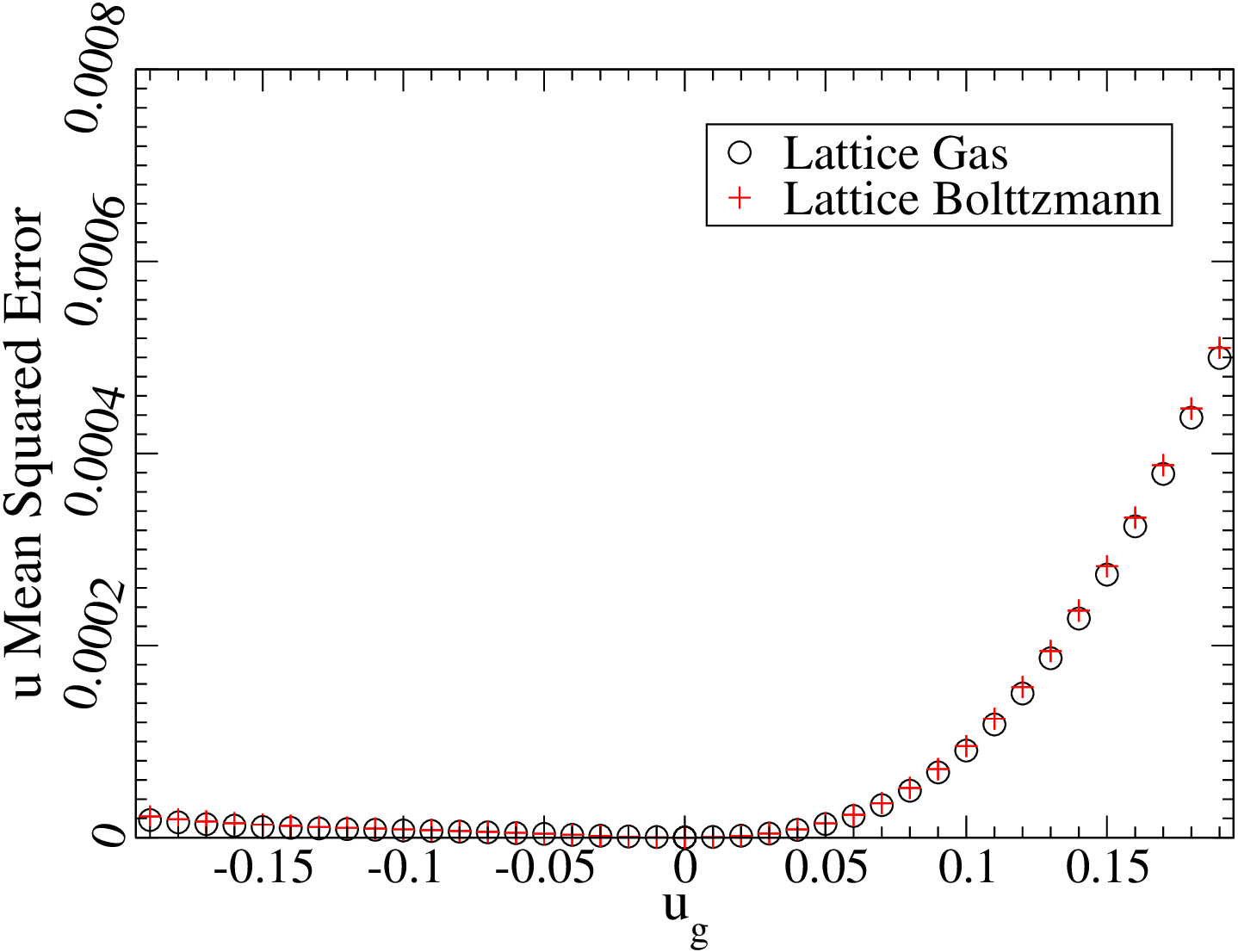}\\(b)
    \caption{Mean Square Error for both the lattice gas and lattice Boltzmann for $\bar{N}$ (a) and $u$ (b). See text for details.}
    \label{fig: InvarianceError}
\end{figure}

A comparison of the $\epsilon$ values for $\bar{N}$ and $u$ for both the lattice gas and lattice Boltzmann algorithms are shown in Fig. \ref{fig: InvarianceError} (a) and \ref{fig: InvarianceError} (b) for various values of $u_g$ ranging from $-0.2$ to $0.2$. The mean squared error for the lattice gas algorithm is on an order of magnitude comparable to that of the lattice Boltzmann algorithm for both $\bar{N}$ and $u$, meaning that the lattice gas and lattice Boltzmann algorithms are approximately equivalent in terms of Galilean invariance. There is a slight deviation observed for higher values of $u_g$, which implies that the lattice gas algorithm is slightly more Galilean invariant than the lattice Boltzmann algorithm, however we are as of yet unsure why this deviation occurs.

\section{Computational Efficiency}

\subsection{Computational Efficiency Results}
The main drawback of Blommel and Wagner's algorithm was its lack of computational efficiency. This caused the method to remain non-competitive when compared to the contemporary fluctuating lattice Boltzmann. Thus, if we are to present our sampling lattice gas as a viable alternative to the fluctuating lattice Boltzmann for fluctuating systems, we must compare the computational efficiency of our algorithm to that of the lattice Boltzmann algorithm.
\begin{figure}
    \centering
     \includegraphics[width=\columnwidth,clip=true]{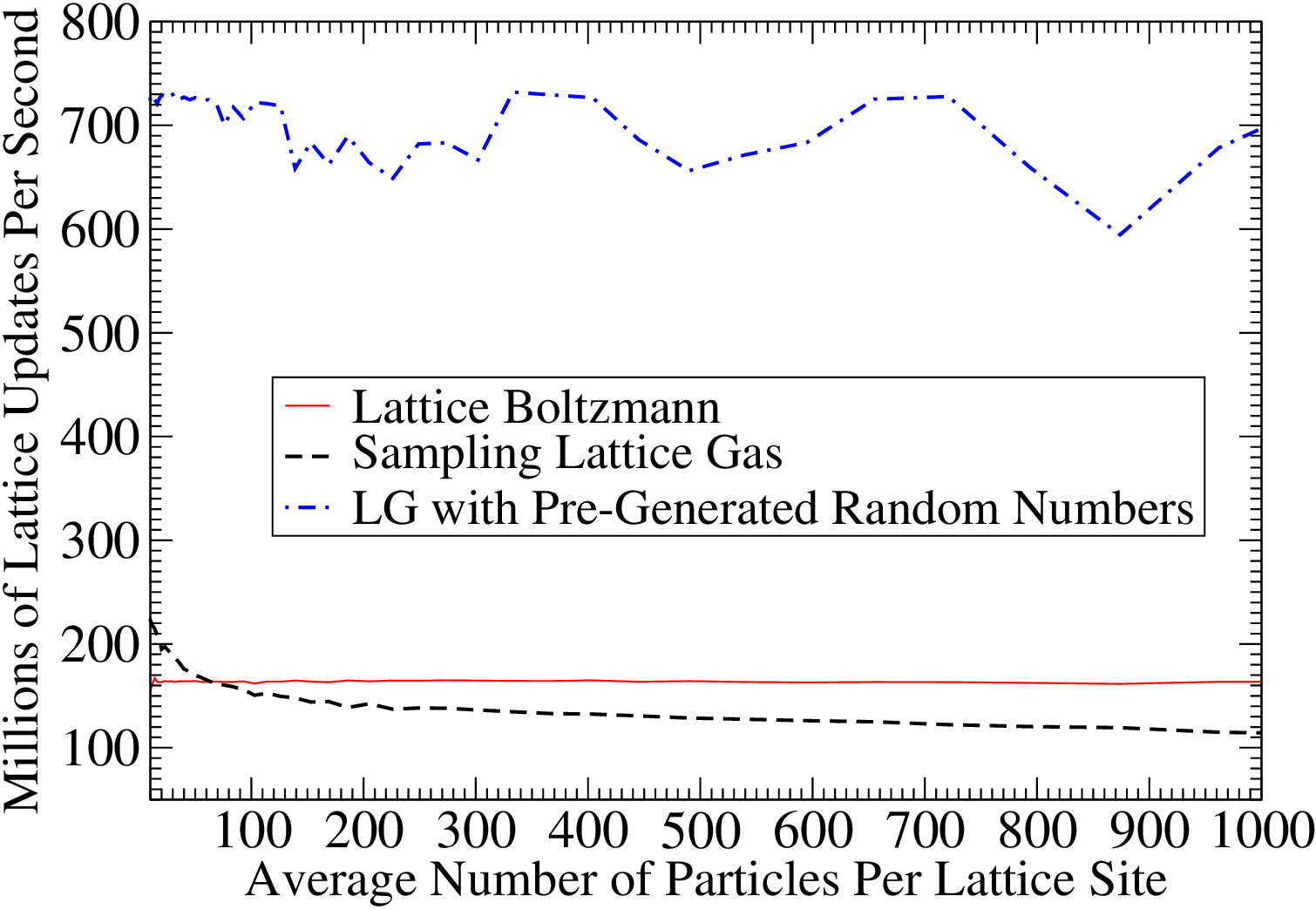} (a)\\
    \includegraphics[width=\columnwidth,clip=true]{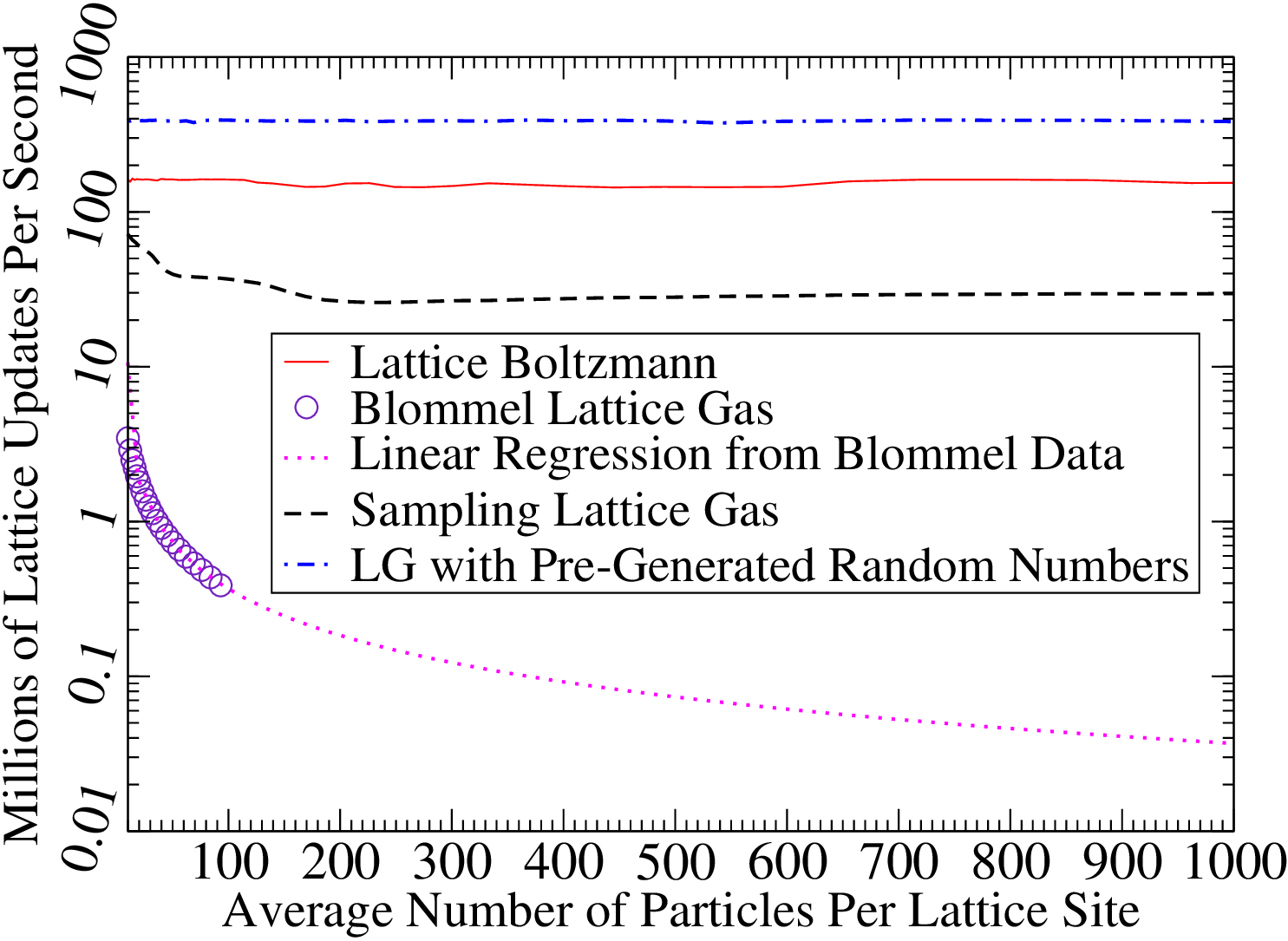} (b)
    \caption{Computational efficiency of the lattice Boltzmann algorithm compared to various lattice gas algorithms for average system densities from $N=10$ to $N=1,000$. (a) compares the lattice Boltzmann algorithm to Blommel and Wagner's algorithm, our sampling lattice gas algorithm, and a second run of our lattice gas algorithm with all randomly sampled numbers pre-generated for $\omega=\omega_{LB}=1$, leaving out Blommel and Wagner's algorithm since it has no true way to reach $\omega=1$. (b) compares the lattice Boltzmann algorithm to our sampling lattice gas algorithm, Blommel and Wagner's algorithm, and a second run of our lattice gas algorithm with all randomly sampled numbers pre-generated for $\omega=\omega_{LB}=0.5$ See text for further details.}
    \label{fig: TimingGraohs}
\end{figure}

Fig. \ref{fig: TimingGraohs} shows the results of the computational efficiency tests for four algorithms: the lattice Boltzmann discussed in Sec. \ref{sec: Boltzmann Algorithm}, Blommel and Wagner's algorithm, a sampling lattice gas lookup table based algorithm discussed in the next section, and a sampling lattice gas with pre-generated random numbers. Specifically, the algorithms are compared based on the number of lattice updates they compute per second, given initial average densities ranging from $N=10$ to $N=1000$ average particles per lattice site. By timing how long it takes for the algorithm to run a particular number of iterations, then converting this time based on the lattice size, we can compare the number of lattice updates directly. For Fig. \ref{fig: TimingGraohs}, these quantities are 10,000 timesteps per run and a lattice size of 100. Due to the high order of magnitude of the number lattice updates per second for the given algorithms, Fig. \ref{fig: TimingGraohs} has units of millions of lattice updates per second, or MLUPS.

It bears noting that the efficiency can vary wildly based on the specs of computer used for simulation. For consistency, all simulations were run on the same Dell XPS laptop with an Intel I5 CPU, and a clock speed of 1.6 GHz. Along with this, because of the way the lookup tables for the sampling lattice gas (which will be discussed further in the next section) work, the computational cost of generating them drops to zero as the simulation time increases, thus we run each lattice gas system for 10,000 timesteps prior to the 10,000 timesteps we time to minimize the cost of this generation. It is also worth noting that the entropic lattice Boltzmann algorithm we are using for comparison is not a fluctuating method, and thus is not truly comparable to the lattice gas system in terms of efficiency. A true fluctuating entropic lattice Boltzmann method would be less efficient than the algorithm used here. Analysis of the local equilibrium $\pi$ ensemble also suggests that for high density systems the distribution becomes approximately normal. Using this approximation to sample, as has been done by Chopard \textit{et al.}~\cite[sec. 5.7]{chopard1998multiparticle, chopardcellular}, may provide even more efficiency for our lattice gas than the explicit distribution sampling methods used here due to the relative efficiency of sampling from a normal distribution.

In Fig. \ref{fig: TimingGraohs} (a), the sampling lattice gas simulates an average of approximately 122.6 MLUPS for densities greater than or equal to 600 average particles per lattice site, and an average of approximately 175.8 MLUPS for densities less than 600 average particles per lattice site. The lattice Boltzmann simulates a relatively consistent average of approximately 162.8 MLUPS for all densities. Thus, for low densities the sampling lattice gas is more efficient than the lattice Boltzmann algorithm, with an average efficiency approximately 7.9\% higher than the lattice Boltzmann. For high density systems, the lattice gas is still on the same order of magnitude as the lattice Boltzmann, with an efficiency loss of approximately 24.7\%.

The $\omega=1$ case shown here does not include Blommel's algorithm, due to the fact that to approximate a particular value of $\omega$ for Blommel's algorithm requires that an associated number of collisions must be done to reach the appropriate number of collided particles on average. The number of collisions $C$ needed for a lattice site with $N$ particles to reach a particular value of $\omega$ is given by,~\cite{seekins2022integer}
\begin{equation}
    C=N\ln(1-\omega).
\end{equation}
In the $\omega=1$ limit, $C$ goes to infinity, and thus $\omega=1$ cannot perfectly be achieved by Blommel and Wagner's algorithm.

In Fig. \ref{fig: TimingGraohs}  (b), the sampling lattice gas simulates an average of approximately 29.2 MLUPS for densities greater than or equal to 600 average particles per lattice site, and an average of approximately 49.7 MLUPS for densities less than 600 average particles per lattice site. As Fig. \ref{fig: TimingGraohs}  (b) represents an under-relaxed system, which requires an additional binomial sampling step at each lattice site for the lattice gas, it follows that the efficiency will be lower. To minimize this inefficiency, we used a binomial sampling method from the GNU Scientific Library (GSL)~\cite{GNU}.

The under-relaxed sampling lattice gas is approximately 4.19 and 3.53 times slower for high and low densities respectively compared to the sampling lattice gas for $\omega=1$. The under-relaxed lattice Boltzmann too is slightly slower than its counterpart at $\omega_{LB}=1$, however this is only by a small margin, simulating an average of approximately 156.5 MLUPS. This represents an efficiency drop of only approximately 4\%, and thus the lattice Boltzmann for an under-relaxed system has an efficiency that is approximately 5.35 and 3.14 times faster than that of the lattice gas under the same circumstances.

Fig. \ref{fig: TimingGraohs} (b) also shows the extent to which our algorithm improves on Blommel's algorithm. The lattice gas in the under-relaxed case is much faster than Blommel's algorithm, including for low densities. The data point for Blommel and Wagner's algorithm with the highest efficiency is already approximately 3.99 times less efficient than the equivalent point for the sampling lattice gas, and the measured data point with the lowest efficiency is approximately 96.2 times less efficient. This trend then continues in the linear regression performed on the measured Blommel data, as the data point with the lowest efficiency of the linear regressed data is approximately 777.3 times less efficient than the equivalent point on the sampling lattice gas.

Finally, we wished to show what the potential maximum increase in efficiency due to the sampling of randomly distributed numbers might be for our sampling lattice gas algorithm. As we are not experts in the efficient random sampling of probability distributions, we find it likely that there are improvements to our algorithm that we could have made but missed. Thus, we pre-generated a set of properly distributed random numbers for each simulation by running a full simulation once, saving all randomly generated numbers for the local equilibrium ensemble and the binomial distribution used to simulate the particle fraction. we then read off those random numbers while timing a second simulation.

In Fig. \ref{fig: TimingGraohs} (a), the sampling lattice gas with pre-generated random numbers had a relatively consistent average for all densities of approximately 708 MLUPS, which is approximately 3.99 times more efficient than the low density lattice gas, and is approximately 4.31 times more efficient than the lattice Boltzmann algorithm. For Fig. \ref{fig: TimingGraohs} (b), the sampling lattice gas with pre-generated random numbers had a relatively consistent average for all densities of approximately 387.1 MLUPS, which is approximately 7.78 times more efficient than the low density lattice gas, and is still approximately 2.47 times more efficient than the lattice Boltzmann algorithm. These results show that there is still a great deal of potential in improving the sampling algorithms used by our lattice gas algorithm, including the potential of becoming even more efficient than the equivalent lattice Boltzmann algorithm.

\subsection{Implementation of the Sampling Lattice Gas Algorithm}

It remains to discuss the exact form of the sampling lattice gas algorithm used for the efficiency tests in the prior section. All our algorithms are written in C, and the full algorithm that we will be discussing here can be found on our GitHub page~\cite{SeekinsHydroMCLG2025Git}.

Because the local equilibrium ensemble out of which we wish to sample is not a well known distribution, using a library such as the GSL in C is non-trivial~\cite{GNU}. Though packages such as the GSL have the ability to sample from arbitrary distributions, the full distribution must be known in advance, which is not the case for our system. This is due to the fact that the $N$ and $J$ values that define the local $\pi$ equilibrium ensemble are not known until after the streaming step of the prior timestep. The ideal solution to this issue would be a library of full distributions pre-calculated for every combination of $N$ and $J$ values, however this method requires either a large amount of memory (scaling with $O(N)$) or to read the distribution from a file for each collision.

Thus, we decided to utilize a dynamic lookup table. This method is similar to what was done in our prior paper on the diffusive sampling lattice gas, only with the added need to manage the $J$ value as well as the $N$ value~\cite{seekins2022integer}. We first initialized the lookup table as a two-dimensional triangular array with $2N-1$ entries in each sub-array, with a likely maximum amount of particles that would be at one lattice site given by three times the average system density. Should a state occur beyond this maximum density, the table would be rebuilt with a higher maximum density in the algorithm. As the system was running, the approximate center point of the cumulative equilibrium distribution ($\sum_{\pi=J} P^0(\pi; N, J)\approx 0.5$) of the local equilibrium ensemble was saved in the lookup table. This then allowed us to then navigate to any part of the local equilibrium ensemble by utilizing the recursive formula from Eq. (\ref{eqn: Pi Dist Recursive}) and its inverse. Because this dynamic lookup table adds entries on the fly, over the course of longer runs generating new table entries will become rare.
\begin{figure}
\begin{algorithm}[H]
\caption{PiGenerate(N,J)}
\label{alg: PiGenerate}
\begin{algorithmic}
    \State int pi=J
    \State double Sum=1
    \State double Log=0
    \While{pi$<$N}
        \State Log=Log+$\ln((\mathrm{N}-\mathrm{pi})(N-\mathrm{pi}-1)/(4(\mathrm{pi}+2)^2+\mathrm{J}^2$))
        \State Sum=Sum+Log
        \State pi=pi+2
    \EndWhile
    \State pi=J
    \State Log=-$\ln$(Sum)
    \State Sum=$\exp$(Log)
    \While{Sum$<$0.5}
        \State Log=Log+$\ln((\mathrm{N}-\mathrm{pi})(\mathrm{N}-\mathrm{pi}-1)/(4(\mathrm{pi}+2)^2+\mathrm{J}^2$))
        \State Sum=Sum+Log
        \State pi=pi+2
    \EndWhile
    \State Save pi as Lookup.PiHalf
    \State Save Log as Lookup.LogHalf
    \State Save Sum as Lookup.SumHalf
\end{algorithmic}
\end{algorithm}
\end{figure}
Generating this lookup table of saved points requires calculating the whole local equilibrium ensemble for $N$ and $J$, as we do not have an explicit formula for the cumulative distribution, and Eq. (\ref{eqn: Pi Dist Recursive}) is not normalized by default. Alg. \ref{alg: PiGenerate} provides a pseudocode representation of the function that generates a lookup table entry for a particular $N$ and $J$ pair.

The natural logarithms are used in Alg. \ref{alg: PiGenerate} due to the fact that for large values of $N$, the unlikelihood of the $\pi$ values at the ends of the distribution leads to floating point errors, something that using the logarithms of these values helps curb. The lookup table itself is, as was described previously, a triangular array with an element for each pair of $N$ and $J$ values. It is initialized as a C structure with three saved values: SumHalf (type double), LogHalf (type double), and PiHalf (type int). This is because to save the closest point of the cumulative local equilibrium ensemble to 0.5 (SumHalf) and be able to iterate up and down the distribution later with Eq. (\ref{eqn: Pi Dist Recursive}) and its inverse, we must know the $\pi$ value at which the SumHalf value occurs, as well as the value of $P^0$(PiHalf$;N,J)$, which we save as its natural logarithm for the reason discussed previously (LogHalf).
\begin{figure}
\begin{algorithm}[H]
\caption{PiLookup(N,J)}
\label{alg: PiLookup}
\begin{algorithmic}
    \If{No Lookup Table exists for this N and J pair}
        \Call{PiGenerate}{N,J}
    \EndIf
    \State Generate (double) $r$, a uniformly distributed random number between 0 and 1
    \State double Sum=Lookup.SumHalf (read from Lookup table)
    \State double Log=Lookup.LogHalf (read from lookup table)
    \State int pi=Lookup.PiHalf (read from the lookup table)
    \If{r$>$Sum}
        \While{Sum$>$r}
            \State Log=Log+$\ln((\mathrm{N}-\mathrm{pi})(\mathrm{N}-\mathrm{pi}-1)/(4(\mathrm{pi}+2)^2+\mathrm{J}^2$))
            \State Sum=Sum+$\exp$(Log)
            \State pi=pi+2
        \EndWhile
    \Else
        \While{Sum$<$r}
            \State pi=pi-2
            \State Log=Log-$\ln((\mathrm{N}-\mathrm{pi})(\mathrm{N}-\mathrm{pi}-1)/(4(\mathrm{pi}+2)^2+\mathrm{J}^2$))
            \State Sum=Sum+$\exp$(Log)
        \EndWhile
    \EndIf
    \State \Return pi.
\end{algorithmic}
\end{algorithm}
\end{figure}
Alg. \ref{alg: PiGenerate} is then called each time a new lookup table element is required. These entries are then used to sample a new value of $\pi$ for the $N$ and $J$ values provided. Alg. \ref{alg: PiLookup} shows how the saved lookup table values are used to efficiently sample a new value of $\pi$ out of our equilibrium ensemble, using a similar process of logarithmic adding to Alg. \ref{alg: PiGenerate}.
\begin{figure}
\begin{algorithm}[H]
\caption{Single Lattice Site Iteration Routine}
\label{alg: Iteration}
\begin{algorithmic}
    \If{$\omega<1$}
        \For{all velocities $v_i$}
            \State int $n^\omega_i$=\Call{GSLBinomial}{$n_i$,$\omega$}
            \State int $n_i^{uncol}=n_i-n_i^\omega$
        \EndFor
        \State int N=$n_{-1}^\omega+n_{0}^\omega+n_{1}^\omega$
        \State int J=$n_{1}^\omega-n_{-1}^\omega$      
        \If{$J\geq0$}
            \State int pi=\Call{PiLookup}{N,J}
            \State $n_{-1}^\omega$=(pi-J)/2
            \State $n_0^\omega$=N-pi
            \State $n_1^\omega$= (pi+J)/2
        \Else 
            \State J=-J;
            \State int $\pi$=\Call{PiLookup}{N,J} 
            \State $n_{-1}^\omega$=(pi+J)/2
            \State $n_0^\omega$=N-pi
            \State $n_1^\omega$= (pi-J)/2
        \EndIf
        \For{all velocities $v_i$}
            \State $n_i=n_i^\omega+n_i^{uncol}$
        \EndFor
    \Else
        \State int N=$n_{-1}+n_0+n_1$
        \State int J=$n_{1}-n_{-1}$      
        \If{$J\geq0$}
            \State int pi=\Call{PiLookup}{N,J}
            \State $n_{-1}$=(pi-J)/2
            \State $n_0$=N-pi
            \State $n_1$= (pi+J)/2
        \Else 
            \State J\=-J;
            \State int pi=\Call{PiLookup}{N,J}
            \State $n_{-1}$=(pi+J)/2
            \State $n_0$=N-pi
            \State $n_1$= (pi-J)/2
        \EndIf
    \EndIf
    \State The particles are now in a redistributed state and are ready to be streamed once all lattice sites are redistributed.
\end{algorithmic}
\end{algorithm}
\end{figure}
Alg. \ref{alg: PiLookup} is used within our full iteration algorithm to sample a new value of $\pi$ out of our equilibrium ensemble. An example of this collision step for a single lattice site can be seen in Alg. \ref{alg: Iteration}. This collision algorithm assumes that the $n_i$ values are stored as an array of integers that has global scope, and accounts for both $\omega=1$ and $\omega<1$ systems. This is a practical implementation of the under-relaxation algorithm discussed in Sec. II A. $\omega$ values greater than one are treated identically to $\omega=1$, as this algorithm does not include overrelaxation.

The function GSLBinomial represents the function from the GSL package that returns a binomially sampled random number if given the number of trials (particles) $N$ and the probability of success (collision fraction) $\omega$~\cite{GNU}. GSLBinomial is not the actual function name, and more implementation is needed to use the GSL sampling method than is included here, but we do not wish to over-complicate this section with a full technical description of its implementation. Please consult the documentation for the GSL for more thorough details. The $n_i^{uncol}$ values, then, represent the total number of particles of each density not selected for collision by the GSLBinomial function.

As the local equilibrium $\pi$ distribution is identical for positive and negative values of $J$, we account for negative values of $J$ in Alg. \ref{alg: Iteration} by flipping the signs on the $J$ values in Eqs. (\ref{eqn: n1 Moment Def}) and (\ref{eqn: nm1 Moment Def}).

Algs. \ref{alg: PiGenerate}, \ref{alg: PiLookup}, and \ref{alg: Iteration} combined represent the full process of collision for our lattice gas algorithm. Alg. \ref{alg: Iteration} is then repeated over all lattice sites, and then the particles are streamed between lattice sites. Neither of these steps are novel, and thus will not be extrapolated upon further here.


\section{Outlook}
Integer lattice gases are a promising simulation method for fluctuating systems, and they can also be directly derivable from Molecular Dynamics simulations via the MDLG method~\cite{parsa2017lattice}. This is why we were motivated to look into them more closely. They also form a physical microsystem that has a Boltzmann limit that can lead to novel insights into developing unconditionally stable lattice Boltzmann methods. Lattice gas methods generally contain additional physical content like correlations, that cannot be captured with current lattice Boltzmann methods, as we saw in section III. Therefore integer lattice gases are an important bridge between molecular representations and their approximation through lattice Boltzmann methods.

There are several remaining open problems when considering the development of the sampling lattice gas algorithm. The most immediately practical of these is the extension to two and three dimensional systems. We are working on this problem and hope to publish a working prototype for this in the near future. We anticipate that this approach will fully recover all of the capabilities of the Blommel and Wagner collision operator.

The main hurdle for this development is the increase in the number of non-conserved moments in higher dimensions. In a D2Q9 system, for example, there are six non-conserved moments, while in a D3Q27 system there are 24 non-conserved moments. The local equilibrium ensemble for the system will, therefore, require us to find the proper distributions out of which to sample these non-conserved moments. We estimate that the extension to a D2Q9 system would be approximately 6 times less efficient for $\omega=1$ systems, and a factor of approximately 3 times less efficient for under-relaxed systems, due to the increased requirement for sampling specific random numbers.

We can also consider extending the velocity space of the system as a whole, which we estimate would have similar increased computational costs relative to the number of velocities added. The more significant hurdle to extending into multi-speed lattices would be identical to that of extending to higher dimensions: finding the proper local equilibrium ensemble out of which to sample.

Overrelaxation is a further open question, crucial for many possible practical applications of the method. We anticipate publishing a promising approach for this for the one dimensional model presented in this paper shortly.

Adding forcing to the collision operator is another important extension of our approach, which would allow the simulation of non-ideal systems. This addition may utilize the coarse-graining of a molecular dynamics simulation, as was done by Parsa and Wagner previously for the ideal lattice gas, to analyze the form of the forcing lattice gas collision operator~\cite{parsa2017lattice}. 

In terms of studying the algorithm itself, the non-uniqueness of the collision operator and the nature of the method's fluctuations in equilibrium and non-equilibrium systems warrant additional attention. Along with these two, the issue of potential non-equilibrium correlations and their effect on relaxation at low densities is another open problem to be considered.

Finally, further optimization of the sampling lattice gas algorithm is needed to truly match the efficiency of or outperform the lattice Boltzmann algorithm for hydrodynamic simulations. Though the lattice Boltzmann algorithm may have the advantage of removing the noise from simulations inherently, those systems that require natural fluctuations would find lattice gas algorithms to be a superior algorithm for practical simulation.

\begin{acknowledgements}
We would like to thank Matteo Lulli for discussing with us the potential influence of correlations on the collision operator, as well as ways to analyze this correlated state more thoroughly for future work. We would also like to thank Victor Ambrus for pointing out the isothermal SOD shock tube solution~\cite{Negro2019Comparison}.
\end{acknowledgements}

\appendix


\bibliographystyle{apsrev4-2}
\bibliography{Bibs/AW,Bibs/MCLG,Bibs/IntegerLG,Bibs/LB,Bibs/SODShockTube}

\end{document}